\documentclass[12pt]{article}
\usepackage{psfig}

\def\lesssim{\mathrel{\mathpalette\vereq<}}
\def\gtrsim{\mathrel{\mathpalette\vereq>}}
\makeatletter
\def\vereq#1#2{
\lower3pt\vbox{\baselineskip1.5pt \lineskip1.5pt
\ialign{$\m@th#1\hfill##\hfil$\crcr#2\crcr\sim\crcr}}}
\makeatother

\begin{document}
\begin{titlepage}
\begin{center}
\hfill    CERN-TH/2001-117\\
~{} \hfill hep-ph/0104312\\

\vskip 1cm

{\large \bf Neutrino Masses and Lepton Flavour Violation in \\
Thick Brane Scenarios}

\vskip 1cm

Gabriela Barenboim,\footnote{gabriela.barenboim@cern.ch}
G. C. Branco,\footnote{gbranco@thwgs.cern.ch and 
gbranco@cfif.ist.utl.pt}$^\P$
Andr\' e de Gouv\^ ea,\footnote{degouvea@mail.cern.ch}
and 
M. N. Rebelo
\footnote{mrebelo@thwgs.cern.ch and
rebelo@cfif.ist.utl.pt}\footnote[5]{On leave of absence from 
Departamento de F\'{\i}sica,  Instituto Superior T\' ecnico \\
Av. Rovisco Pais, P-1049-001, Lisboa,
Portugal.} 
\vskip 0.05in

{\em Theory Division, CERN, CH-1211 Geneva 23, Switzerland.}
\end{center}

\vskip 3cm

\begin{abstract}

We address the issue of lepton flavour violation and neutrino masses
in the ``fat-brane'' paradigm, where flavour changing processes are suppressed by 
localising different fermion field wave-functions at different positions 
(in the extra dimensions) in a thick brane.
We study the consequences of suppressing lepton number violating charged lepton decays
within this scenario for lepton masses and mixing angles. In particular, we find
that charged lepton mass matrices are constrained to be 
quasi-diagonal. We further consider whether 
the same paradigm can be used to naturally explain small Dirac neutrino 
masses by considering the existence of three right-handed neutrinos 
{\sl in the brane}, and discuss the requirements to obtain phenomenologically 
viable neutrino masses and mixing angles. 
Finally, we examine models where neutrinos obtain a small Majorana mass by breaking 
lepton number in a far away brane and show that, if the fat-brane paradigm is the
solution to the absence of lepton number violating charged lepton decays, such models
predict, in the absence of flavour symmetries,
that charged lepton flavour violation will be observed in the next round
of rare muon/tau decay experiments.  

\end{abstract}

\end{titlepage}

\newpage
\setcounter{footnote}{0}
\setcounter{equation}{0}
\section{Introduction}

A few years ago, a new paradigm for addressing the gauge hierarchy problem 
was proposed \cite{extradims}. In the so-called ``large extra dimensions scenario,''
(ADD scenario)
the explanation for the large discrepancy between the weak scale and the Planck
scale is that the fundamental scale of gravity is actually close to the
weak scale, and that gravity seems weak to us because it propagates in more, compact,
dimensions, while Standard Model fields are bound to four-dimensional subspaces 
embedded into the higher dimensional world.    

One of the challenges for the ADD scenario is solving the
well known problems of proton stability, absence of flavour changing neutral current
processes, etc. 
The situation is particularly problematic because one has to consider 
nonrenormalizable operators which are only suppressed by powers of the ``quantum
gravity scale'' and whose coefficients are completely unknown since they parametrise
our ignorance concerning the physics above the ultraviolet cutoff. It is
important to ask whether there are ``low energy'' mechanisms guaranteeing 
that most of these dangerous terms are appropriately suppressed, independently of 
the unknown beyond-quantum-field-theory physics. 

A novel solution to the higher-dimensional flavour problems was proposed by 
Arkani-Hamed and Schmaltz \cite{AS} (AS scenario). 
They argue that if we live in a ``fat'' 
four-dimensional spacetime (fat brane), it is possible to effectively suppress
unwanted higher dimensional operators by localising different fermion fields
to different coordinates in the extra dimensions. As was also pointed out in 
\cite{AS}, and further explored in \cite{MS,BdGR}, this solution also provides an 
interesting explanation for the hierarchy of the quark masses and the values of the 
CKM mixing angles and CP-violating phase. It should be pointed that the AS scenario 
leads to startling consequences \cite{AGS}, which may be observed at 
next-generation collider experiments. 

In this paper, we address the issue of lepton flavour violating phenomena in 
charged lepton decays
and neutrino masses in the AS scenario. The AS scenario is natural for solving
flavour changing problems, and indeed we find that it can very efficiently suppress
charged lepton flavour violating decays such as $\mu,\tau\rightarrow l\gamma, lll$.
One of the issues we explore here is whether suppressing such processes imposes any
interesting constraint on lepton masses and mixing angles. We find, for example,
that the charged lepton mass matrix is forced to be quasi-diagonal, and that almost
all the mixing in the lepton sector {\sl must} come from the neutral sector.
Such a situation is certainly not observed in the quark sector (for an example, 
see \cite{BdGR}), and seems to already constrain some extra-dimensional neutrino 
mass scenarios.

The AS scenario also provides a natural mechanism for
obtaining small Dirac neutrino masses, which has not yet been 
explored in the literature. It is important, therefore, to study whether
similar success is obtained in explaining, for example, the 
large leptonic mixing angle which is required by neutrino oscillation solutions 
to the atmospheric neutrino puzzle \cite{neutrinos}. 
We will show that in order to obtain large mixing angles, separations between
different leptonic fields have to be closely related. The situation appears to be 
rather finely tuned when the flavour constraints mentioned above are also imposed.

The paper is organised as follows. In Sec.~2, we address the problem of flavour 
changing neutral currents in the
ADD paradigm and how it is solved in the AS scenario. We concentrate on 
flavour violating charged lepton decays. In Sec.~3, we address the issue of 
lepton masses and mixing angles in a straightforward extension of the AS scenario, 
namely, adding right-handed neutrinos to the fat brane. 
We first discuss how ``naturally''
the AS scenario can explain the large hierarchies observed in the quark and fermion
sectors, and what is required in order to obtain large mixing angles. We then analyse 
two-family models in great detail, concentrating on the constraints that must be
met in order to obtain large mixing angles. Finally, we discuss solutions to the
neutrino puzzles in the case of three generations. In Sec.~4, we discuss 
a mechanism for obtaining small neutrino masses without right-handed neutrinos. Here,
constraints from the charged lepton sector severely restrict solutions to the
neutrino puzzles. Our conclusions are presented in Sec.~5.

\setcounter{footnote}{0}
\setcounter{equation}{0}
\section{Lepton Flavour Violation and Large Extra Dimensions}

It is a general principle of effective field theories that all operators
consistent with the exact symmetries of Nature are present, including
nonrenormalizable, ``irrelevant'' operators. These operators are understood
as being suppressed by the large energy scale above which the effective 
field theory is no longer applicable. These include, for example, 
$(1/\Lambda^2) QQQL$,\footnote{We use the standard notation where $Q$ and $L$ refer,
respectively, to quark and lepton $SU(2)$ doublets, while $U$, $D$, and $E$ denote,
respectively, up-type antiquark, down-type antiquark, and antilepton $SU(2)$ singlets.
$H$ is the $SU(2)$ Higgs boson doublet.} 
which mediates proton decay and $(1/\Lambda) LHLH$, which
breaks lepton number and gives the neutrinos a Majorana mass after electroweak
symmetry breaking. It is also well known that $\Lambda$ is constrained to
be at least of the order of the grand unification scale 
($M_{GUT}\simeq 10^{16}$~GeV) by the
fact that the proton lifetime is larger than $\sim 3\times 10^{33}$~years 
\cite{p_decay}. On the other hand,
values of $\Lambda\sim 10^{9 - 14}$~GeV are required in order to correctly
account for the atmospheric neutrino puzzle via neutrino oscillations. 

In theories where the ``quantum gravity'' scale is small (a few TeV), one is 
faced with the challenge of explaining why, even though $\Lambda\sim 1$~TeV,
phenomena which violate the conservation of ``accidental'' global 
symmetries have not been
copiously observed. 
The two irrelevant operators mentioned above can be 
eliminated by assuming, for example, that (gauged) $B-L$ is a true 
symmetry of Nature  (probably weakly broken, if one wants to explain 
the matter--antimatter asymmetry of the universe), but other 
dangerous operators still remain. These include operators that violate
individual flavour number conservation and mediate phenomena such as
meson--antimeson mixing and flavour changing neutral currents.

One intriguing solution to these problems was proposed 
by Arkani-Hamed and Schmaltz \cite{AS}. They postulate that the standard 
model fields are constrained to a ``fat brane,'' which is infinite in 
three space directions and occupies a finite volume in $n$ extra, orthogonal, 
compact dimensions. It is
also postulated that gauge fields and the Higgs scalar can freely propagate
in the entire volume of the fat brane, but fermions are confined to 
specific ``points.'' Finally, if one assumes that 
different fermion fields are located at {\sl different} points in the extra 
dimensions, it is easy to see that some operators are
``forbidden'' by locality, {\it e.g.,}\/ an operator such as $(1/\Lambda^2) QQQL$
can be very efficiently suppressed simply because $Q$'s and $L$'s live in
different worlds, and don't ``see'' one another \cite{AS}.

In this section we discuss, within the scenario described above 
(AS scenario), what 
requirements are imposed on the distances between different leptonic fields
in order to account for the (negative) experimental searches for charged
lepton decays which violate lepton flavour number. Here we will concentrate on
the lepton flavour violating muon decays
\begin{equation}
\mu\rightarrow e\gamma, \hspace{2cm} \mu\rightarrow eee,
\end{equation} 
and tau decays\footnote{We do not include the decays 
$\tau\rightarrow\mu ee,~e\mu\mu$ in our discussion.
The reason for this is that
the constraints imposed by these channels
are far weaker than those imposed by $\mu\rightarrow e\gamma,~eee$, 
as will become clear shortly.}
\begin{equation}
\tau\rightarrow\mu\gamma, \hspace{1cm} \tau\rightarrow e \gamma, 
\hspace{1cm} \tau\rightarrow\mu\mu\mu, \hspace{1cm} \tau\rightarrow eee. 
\end{equation}

The process $l^k\rightarrow l^l\gamma$ ($k,l=\tau,\mu,e$) is mediated by
the effective Lagrangian
\begin{equation}
-{\cal L}_{\l^k\rightarrow l^l\gamma}=
\frac{a_R v}{\Lambda^2}\left(\bar{l^k_L}\sigma_{\mu\nu}l^l_R
F^{\mu\nu}\right) +
\frac{a_L v}{\Lambda^2}\left(\bar{l^k_R}\sigma_{\mu\nu}l^l_L
F^{\mu\nu}\right) +
h.c.,
\label{l-lgamma}
\end{equation}
after electroweak symmetry breaking. Here the subscript $L,R$ refers to the 
chirality, $F^{\mu\nu}$ is the electromagnetic field strength, 
$v$ is the Higgs vacuum expectation value, 
and $a_{R,L}$ are dimensionless couplings.

The process $l^k\rightarrow l^ll^ll^l$ is mediated by the effective Lagrangian (after
Fiertz rearrangements)
\begin{eqnarray}
-{\cal L}_{l^k\rightarrow l^ll^ll^l} & = & 
-{\cal L}_{\l^k\rightarrow l^l\gamma}+ \nonumber \\
&& +\frac{1}{\Lambda^2}\left[b_1 (\bar{l^k_R}l^l_L)(\bar{l^l_R}l^l_L)   
+ b_2 (\bar{l^k_L}l^l_R)(\bar{l^l_L}l^l_R)+\right.  \nonumber \\
&& +  b_3 (\bar{l^k_R}\gamma_{\mu}l^l_R)(\bar{l^l_R}\gamma^{\mu}l^l_R) + 
b_4 (\bar{l^k_L}\gamma_{\mu}l^l_L)(\bar{l^k_L}\gamma^{\mu}l^l_L)+ 
\nonumber \\
&& \left. +  b_5 (\bar{l^k_R}\gamma_{\mu}l^l_R)(\bar{l^l_L}\gamma^{\mu}l^l_L) +
b_6 (\bar{l^k_L}\gamma_{\mu}l^l_L)(\bar{l^l_R}\gamma^{\mu}l^l_R)+ h.c.\right], 
\label{l-3l}
\end{eqnarray}
where $b$'s are dimensionless couplings.
Eq.~(\ref{l-lgamma}) mediates $l^k\rightarrow l^ll^ll^l$ 
by attaching an $l^l\bar{l^l}$ pair to the photon leg. 

In the AS scenario, one can relate the constants $a,b$ to the distance
between different leptonic fields and to higher dimensional couplings by
integrating out the extra dimensions. One obtains, for example,
\begin{equation}
a_R\propto\alpha_R\int{\rm d}y e^{-\mu^2(y-l_{k})^2}e^{-\mu^2(y-e_{l})^2}
\propto \alpha_R e^{-\frac{1}{2}\mu^2(l_{k}-e_{l})^2}
\end{equation}
and
\begin{equation}
b_4\propto\beta_4\int{\rm d}y e^{-\mu^2(y-l_{k})^2}\left(e^{-\mu^2(y-l_{l})^2}
\right)^3
\propto \beta_4 e^{-\frac{3}{4}\mu^2(l_{k}-l_{l})^2},
\end{equation}
assuming one extra dimension and defining the higher-dimensional wave-function
of the fermions to be Gaussians of width $\mu^{-1}/\sqrt{2}$ centred at position  
$l_{k,l}$ and $e_{k,l}$ for $l_L^{k,l}$ and $l_R^{k,l}$, respectively. The 
couplings $\alpha,\beta$ are dimensionless numbers, which are assumed to be of 
order one, since there is, {\it a priori}\/ no reason for them to be either very 
small or very large. All $\alpha$ and $\beta$ will be, henceforth, set to 
unity unless otherwise noted.

From Eqs.~(\ref{l-lgamma},\ref{l-3l}) and assuming the AS scenario, it is 
easy to compute the branching ratios for the various lepton decays (see, for
example, \cite{KO}) as a function of $\Lambda$ (the ``quantum gravity'' scale)
and the distance between different lepton fields.  

\begin{figure}
    \centerline{
    \psfig{file=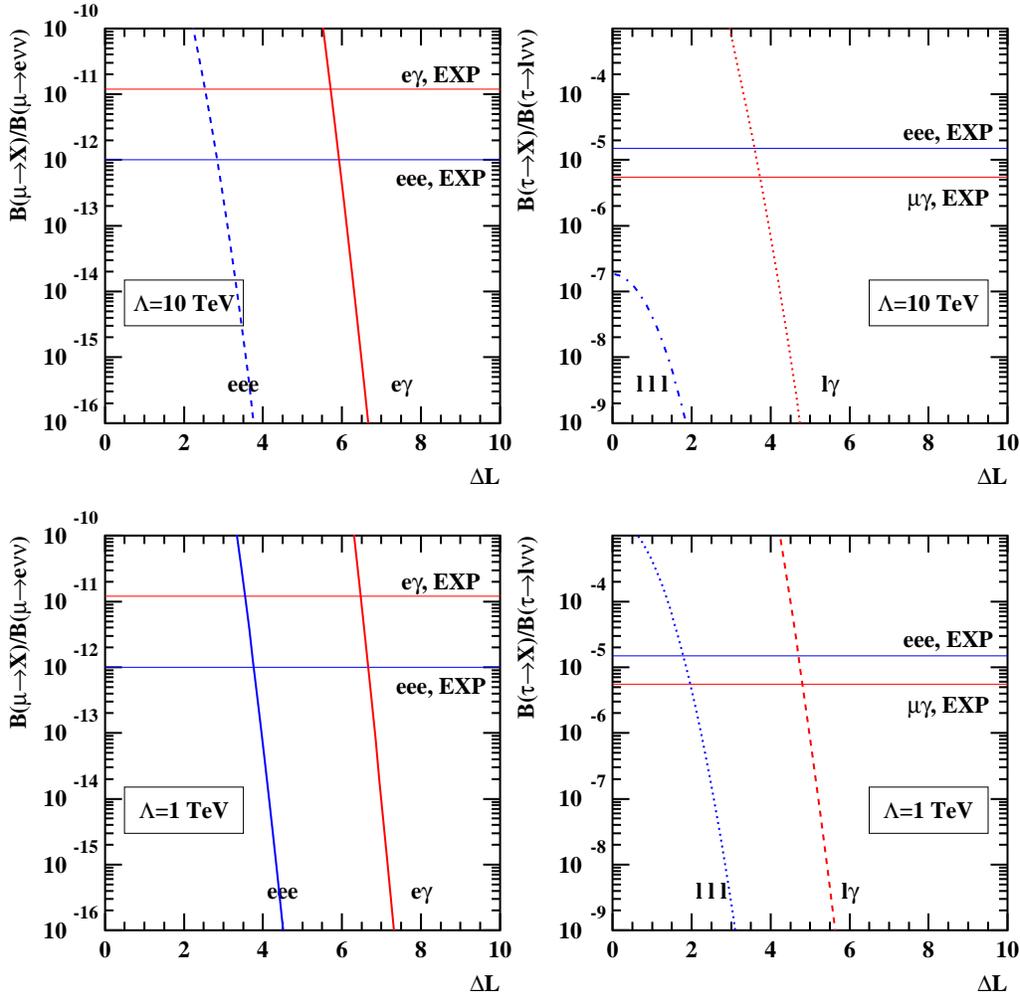,width=1\columnwidth}}
    \caption{Branching ratio for 
$\mu\rightarrow X$ (right) and $\tau\rightarrow X$, 
(left) normalised by the branching ratio for $\mu\rightarrow e\nu\bar{\nu}$
and $\tau\rightarrow l\nu\bar{\nu}$ respectively, as a function of $\Delta L$, 
for two values of the quantum-gravity scale, $\Lambda =10$~GeV (top) and 1~GeV 
(bottom).
See text for details. The horizontal lines indicate the current experimental
constraints from searches for rare muon and tau decays \cite{PDG}. In the case
of tau decays, only the most and least stringent bounds are shown.}
    \label{fig:mu-egamma}
\end{figure}

Fig.~\ref{fig:mu-egamma} depicts the branching ratio for 
$\mu\rightarrow e\gamma$ (right) and $\tau\rightarrow l\gamma$, $l=e,\mu$ 
(left) normalised by the branching ratio for $\mu\rightarrow e\nu\bar{\nu}$ 
and $\tau\rightarrow l\nu\bar{\nu}$ respectively, as a function of $\Delta L$. 
Two different values of the cutoff, $\Lambda=10$~TeV and 1~TeV, are shown 
(top and bottom, respectively). Here, $\Delta L$ is either 
$|l_{\mu}-e_{e}|$ or $|e_{\mu}-l_{e}|$ for the muon decay and 
$|l_{\tau}-e_{l}|$ 
or $|e_{\tau}-l_{l}|$ for the tau decay, in units of $\mu^{-1}$.

Fig.~\ref{fig:mu-egamma} also depicts the branching ratio for 
$\mu\rightarrow eee$ (right) and $\tau\rightarrow lll$, $l=e,\mu$ 
(left) normalised by the branching ratio for $\mu\rightarrow e\nu\bar{\nu}$ 
and $\tau\rightarrow l\nu\bar{\nu}$ respectively, as a function of $\Delta L$. 
Here, only nonzero $b_3$ and $b_4$ coefficients were considered. 
The reason for this is simple: given the requirements imposed by the 
non-observation of $\mu \rightarrow e\gamma$ and $\tau\rightarrow l\gamma$,
the contribution of $a_{R,L},b_{1,2,5,6}$ are strongly suppressed, while
$b_{3,4}$ are, a priori, unconstrained. Therefore, in this case, 
$\Delta L$ is either 
$|l_{\mu}-l_{e}|$ or $|e_{\mu}-e_{e}|$ for the muon decay and 
$|l_{\tau}-l_{l}|$ or $|e_{\tau}-e_{l}|$ for the tau decay, in units of 
$\mu^{-1}$.

Some comments are in order. First of all, it must be said that the AS 
scenario, as expected, is very efficient when it comes to suppressing
rare lepton decays. Even for $\Lambda=1$~TeV, the most stringent
requirement (from $\mu\rightarrow e\gamma$) is that $|l_{\mu}-e_e|, 
|l_e-e_{\mu}|\gtrsim 6.5\mu^{-1}$ 
(proton decay constraints require distances between quarks and lepton 
of order $10\mu^{-1}$ \cite{AS}). Second, one may wonder how
strict the constraints are since, after all, there are order one coefficients
floating about. It is easy to see that things do not change significantly if these
coefficients are allowed to vary. 
For example, varying $\alpha_R$ from 0.3 to 3,
the lower bound on $|\l_{\tau}-e_{\mu}|$ from $\tau\rightarrow\mu\gamma$ and
$\Lambda=10$~TeV changes by $17\%$.
A remaining question is what do these bounds imply for 
lepton masses and mixing angles. This will be addressed in the next sections. 

Before proceeding, however, it is worthwhile mentioning at this point that other
lepton flavour violating decays, such as $K\rightarrow \mu e$,
hadronic tau decays and $\mu$--$e$ conversion in nuclei\footnote{The Lagrangian
Eq.~(\ref{l-3l}) mediates $\mu$--$e$ conversion in nuclei at higher order in
QED (in the case of the magnetic moment operators) and at higher loop level and
higher order in QED in the case of the four-fermion operators. The constraints
imposed on the operators are not as strict as the ones from lepton flavour
violating charged 
lepton decays.} have not been 
considered, since they can be safely suppressed by assuming that quarks and 
leptons are well separated (which is generically required in order to 
suppress proton decay), and that the distance constraints obtained in these 
cases do not play a significant role when it comes to addressing lepton 
masses and mixing angles.  

\setcounter{equation}{0}
\setcounter{footnote}{0}
\section{Right-Handed Neutrinos in the Brane}

Another intriguing possibility is using the AS scenario to explain 
the current hierarchy of fermion masses and mixing angles \cite{AS}. Indeed,
it has been shown that localising different quark fields to different points
in the extra dimensions is an efficient and ``natural'' way of explaining not
only all quark masses and mixing angles \cite{MS}, but also the observed
CP violation in the quark sector \cite{BdGR}.   

In this section, we study whether the same idea can be used to explain lepton
masses. Charged lepton masses were already discussed in the context of massless
neutrinos \cite{MS}, but no similar study has been performed for 
neutrino masses.

The idea is quite simple. We assume the existence of three standard model
singlet fermions,
which we refer to as right-handed neutrinos ($N$), and assume that these are also
localised at different points in the extra dimensions. If this is the case,
neutrinos acquire ordinary Dirac masses after electroweak symmetry breaking, 
exactly like the quarks and charged leptons. In the AS scenario, the explanation 
for the smallness of the neutrino masses is straightforward: neutrino masses
are tiny because $N$'s and $L$'s are separated by distances which are only 
slightly larger (as will become clear later) than, for example, the $Q$'s and 
$D$'s.\footnote{Another possibility for obtaining naturally small Dirac neutrino
 masses is to assume that the neutrinos couple to right-handed neutrinos
 which live in the bulk \cite{bulk}. Lately, this possibility has received lots 
 of attention \cite{bulk_N}, and will not be discussed in 
 this paper.}
It will turn out, as we will discuss in detail later, that the most serious
challenge for the AS scenario with right-handed neutrinos in the brane will be
to naturally account for the large mixing angles in the lepton sector which are
required in order to solve the neutrino puzzles \cite{neutrinos}.

Before proceeding, it is worthwhile to digress a little on the philosophy 
behind the AS scenario when it comes to explaining fermion masses and mixing angles.
Perhaps the most attractive, and certainly the novel idea behind the AS scenario
is the possibility of solving the flavour problem without postulating new,
horizontal symmetries. Instead, it is simply assumed that the fermions are
separated by distances of a few (in units of the width of the fermionic
higher dimensional wave-function $\mu^{-1}$), 
which could be chosen at random. All the work
is done by the Gaussian suppression of the effective four dimensional Yukawa
coupling (as discussed in the previous section. See also \cite{AS,MS}):
\begin{equation}
\lambda\propto e^{-\frac{1}{2}\mu^2(f_i-f_j)^2},
\end{equation} 
where $f_i$ are the positions of the different fermion fields ($Q,U,D,L,E,N$). 

\begin{figure}
    \centerline{
    \psfig{file=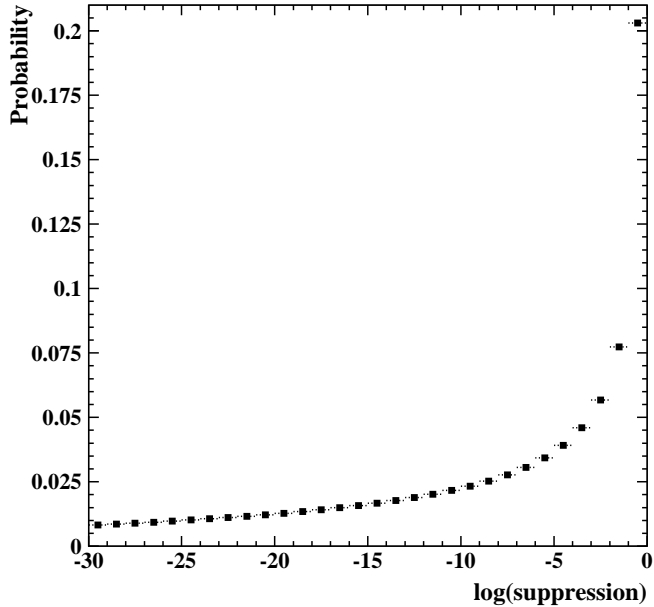,width=0.6\columnwidth}}
    \caption{The probability for obtaining a suppression of the higher
dimensional Yukawa coupling of a given order of magnitude. See text for details. }
    \label{fig:random}
\end{figure}

In order to further quantify the ``naturalness'' concept, let us assume that, say, 
a $Q$ and a $U$ fields are localised in one extra dimension, of size 
$L \equiv 20\mu^{-1}$ (see \cite{MS} for a discussion on the expected size of
$L\mu$). If the positions are chosen completely at random\footnote{It has recently
been discussed in \cite{NuSh} that the relative distance between different fermion
fields may be significantly altered by gauge interactions. We will disregard this
effect for the sake of simplicity.} it is easy to show that the 
probability of finding $|q-u|\mu$ between some value $\delta$ and $\delta + {\rm d}
\delta$ is given by $P(\delta){\rm d}\delta$, where
\begin{equation}
P(\delta)= 0.1 - .005\delta.
\end{equation}
It is straightforward to compute the probability that the suppression due to 
the small overlap between the two extra dimensional wave-functions is of a 
given order of magnitude. Fig.~\ref{fig:random} depicts the probability that
the suppression factor is in some order of magnitude bin. It should be noted
that while the figure only depicts suppression factors larger than $10^{-30}$,
the probability that the suppression factor is smaller than $10^{-40}$ is almost
10\%!

As expected, very tiny suppression factors can be obtained, even if all fields are
constrained to lie relatively close to one another. An important fact is that,
while random choices for the positions (in one extra dimension) seem to favour
effective Yukawa coupling which are suppressed by an order of magnitude with
respect to their higher dimensional counterparts, the probability of obtaining 
a large spread of suppression factors is quite large. It is this property which
renders the AS solution to the fermion mass hierarchy attractive -- random choices
for the distances seem to provide a large spread for values of the Yukawa 
couplings, and therefore the observed hierarchies are naturally explained.

One can further pursue this logic and see what are the consequences for the 
mixing angles. We use a toy model of only two families of up-type quarks as an 
example, but it will become clear in due time that it is easily generalised to
include three families and down-type quarks.

The mass matrix can be written as
\begin{equation}
M_U=\rho \left(\matrix{a & b \cr c & d}\right),
\label{2x2_matrix}
\end{equation}    
where $\rho$ is roughly the top quark mass and $a,b,c,d\leq 1$ are the suppression
factors. The message from Fig.~\ref{fig:random} is that, ``normally'' one expects
$a\gg b\gg c\gg d$ (for example). If this is the case, one can trivially show 
that the matrix that diagonalizes $M_UM_U^{\dagger}$ is parametrised by one 
mixing angle, $\theta$, such that $\cos 2\theta=\pm1\mp {\cal O}(c^2/a^2,cbd/a^3)$, 
{\it i.e.,}\/ $\theta$ is either close to 0 or $\pi/2$, such that
the mixing is quite suppressed. 

While this situation fits very nicely the quark masses and
mixing angles, it poses serious problems for leptons, given that, according to the 
neutrino oscillation solution to the neutrino puzzles, at least one of the
leptonic mixing angles 
is ``large.'' This is in sharp contrast to standard four dimensional approaches
to the fermion mass puzzles. There, quark masses require a significant amount
of underlying structure  in order for one to understand the hierarchy of masses
and the small mixing angles, while it has been argued that neutrino masses can
be obtained from operators with random order one couplings \cite{anarchy}. 
The same is true in the case of the ``simplest'' neutrino mass model 
\cite{minimalistic}, where neutrino masses are obtained via the introduction of
one single right-handed neutrino and the use of $1/M_{\rm Planck}$ operators. 

In order to obtain large mixing angles, it is imperative that some elements 
of the mass matrices are of the same order. Using Eq.~(\ref{2x2_matrix}) as an
example, it is again simple to show that, if one wants $\theta\simeq\pi/4$,
$a\simeq c$ (for example) is required. 

The challenge is, therefore, to obtain a mechanism satisfying this condition.
Flavour symmetries, for example, would certainly do the job, but they would
spoil the most attractive feature of the AS scenario. A different option, which
still maintains the extra-dimensional nature of the AS scenario, is to assume
that, as a consequence of the underlying localising mechanism, some fields are
positioned on top of each other, and that some distances are, therefore,
``naturally'' very similar.

With this in mind, we proceed to discuss how realistic lepton masses and 
mixing angles can be 
obtained in the AS scenario. We first discuss the simpler two family case, and
further address the issue of how ``finely tuned'' certain parameters have to be
in order to obtain large mixing. Later, we discuss solutions for
the more involved case of three families.  
 
\subsection{Two Neutrino Families}

\begin{figure}
    \centerline{
    \psfig{file=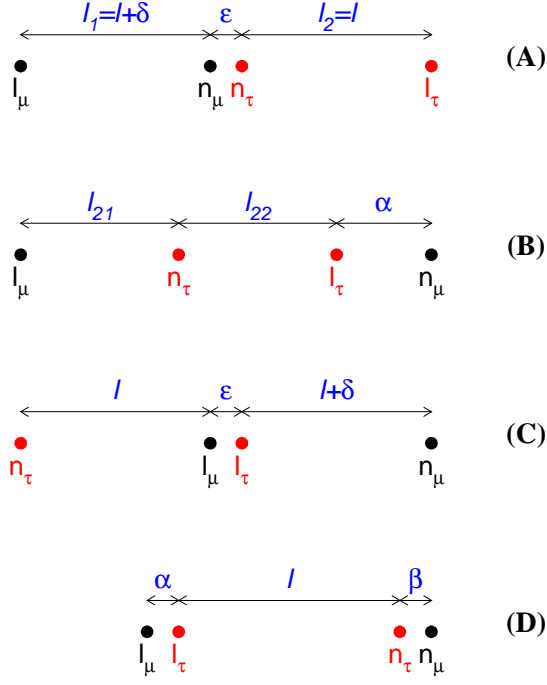,width=0.7\columnwidth}}
    \caption{Different two family configurations which potentially lead to maximal
mixing. }
    \label{geoms}
\end{figure}

In this subsection we address in detail the case of only the two ``heavy''
lepton families, 
concentrating on the observed large mixing which seems to be required
in the $\mu-\tau$ sector according to neutrino oscillation
solutions to the atmospheric neutrino puzzle. It proves convenient to
consider separately the different configurations which potentially yield maximal
mixing in order to understand what constraints over the space of locations
have to be met in order to obtain the appropriate mass-squared difference and mixing
angle. These configurations are depicted in Fig.~\ref{geoms}.

Two important comments concerning our discussion are in order. First, we assume that
the order one coefficients which accompany the exponential suppressions (the higher
dimensional Yukawa couplings) play no significant role, and set all of them to unity.
We comment on the relevancy of this approximation later. 
Second, we impose that there is no significant mixing
coming from the charged lepton sector. It turns out
that, specially when considering only the $\mu-\tau$ sector, the non-observation 
of $\tau\rightarrow\mu\gamma$ combined with the requirement that the obtained 
$\tau$ mass agrees with the experimental value automatically forces the mixing
in the charged lepton sector to be small. Note that the constraint of
small mixing in the charged lepton sector will be considered {\it a posteriori},
and its consequences analysed for each of the different configurations.

By imposing the $\tau\rightarrow\mu\gamma$ constraint as discussed in the previous
section, one obtains $|l_{\mu}-e_{\tau}|, |l_{\tau}-
e_{\mu}|>3.73\mu^{-1}$~$(4.09\mu^{-1})$ for $\Lambda=10$~TeV (5~TeV), 
which in turn means that
the off-diagonal elements of the (two generations) Dirac mass matrix for the
charged leptons are less than (approximately) $\rho e^{-0.5 (4)^2}= 88$~MeV
for $\rho=1.5\times m_t$, where $m_t\equiv 166$~GeV is the top quark mass (for
details concerning the choice of $\rho$, see \cite{MS}). 
This, in turn, implies
that one of the diagonal elements is {\sl necessarily} of the order of the tau mass,
while the other elements are at most of the order of the muon mass. In summary,
\begin{equation}
m_{l} \simeq m_{\tau}\left(\matrix{ \epsilon(1+bc\epsilon) &  b\epsilon
\cr c \epsilon & 1-\frac{\epsilon^2}{2}(b^2+c^2)}\right),
\label{2gen_charged}
\end{equation}
where $b,c$ are numbers which are at most order one, $\epsilon\equiv 
m_{\mu}/m_{\tau}$, and $m_{\mu,\tau}$ are the muon and tau masses, respectively.
Note that Eq.~(\ref{2gen_charged}) is written in such a way that its eigenvalues are
$m_{\mu}$ and $m_{\tau}$ up to $O(\epsilon^2)$. 
In this case, it is easy to estimate the mixing angle, namely,
\begin{equation}
\sin^2\theta \simeq b^2 \frac{m_{\mu}^2}{m_{\tau}^2}\lesssim 10^{-3},
\end{equation}
which is quite small.

A fact that will prove very important in the future is that, 
even though there is no direct significant limit on 
$|l_{\mu}-l_{\tau}|$ from $\tau\rightarrow \mu\mu\mu$ decays (see 
Fig.~\ref{fig:mu-egamma}), it is geometrically constrained because
$|l_{\tau}-e_{\tau}|<|l_{\mu}-e_{\tau}|$. It is easy to show that 
$|l_{\mu}-l_{\tau}|>|l_{\mu}-e_{\tau}|-|l_{\tau}-e_{\tau}|\simeq 
(4.0-3.14)\mu^{-1}$,\footnote{Note that $m_{\tau}=1777$~MeV corresponds to a distance of
$3.14\mu^{-1}$, and $m_{\mu}=105.7$~MeV to a distance of
$3.94\mu^{-1}$.}
independent of the number of extra dimensions.
Therefore, as a guarantee that there are no large flavour changing 
effects in the charged lepton sector, a good rule of thumb is to keep
$|l_{\mu}-l_{\tau}|\gtrsim 1\mu^{-1}$, in which case the mixing in the charged lepton 
sector is necessarily small. 

{\it Case A} -- Fig.~\ref{geoms}(A) leads to the neutrino Dirac mass matrix
\begin{equation}
m_{\nu}=\rho \left(\matrix{e^{-\frac{\mu^2}{2}(l+\delta)^2} & 
e^{-\frac{\mu^2}{2}(l+\delta+\epsilon)^2} 
\cr e^{-\frac{\mu^2}{2}(l+\epsilon)^2} & e^{-\frac{\mu^2}{2}l^2}}\right),
\label{case1}
\end{equation} 
where $l\equiv l_2$ and $\delta\equiv l_1-l_2$. We restrict ourselves to $\delta>0$, 
without loss of generality. As mentioned before, we will fix $\rho=1.5\times 
166$~GeV.

It should be noted that the interchange of $n_{\mu}$ with $n_{\tau}$ 
(see Fig.~\ref{geoms}) corresponds to a weak basis transformation which 
interchanges the two columns of $m_{\nu}$ but leaves $m_{\nu}m_{\nu}^{\dagger}$
unchanged. Note also that the configuration depicted in 
Fig.~\ref{geoms}(A) leads automatically to very small mixing in the charged 
lepton sector, once the appropriate sizes for the neutrino masses are imposed.  

Diagonalizing $m_{\nu}m_{\nu}^{\dagger}$ starting from Eq.~(\ref{case1}) one
obtains the mixing angle $\theta$, which, to a very good approximation 
(assuming $\delta,\epsilon\ll l$), is given by
\begin{equation}
\tan 2\theta \simeq \frac{2N_1(1+N_2)}{(D_1-1)(D_2+1)},
\label{tan1}
\end{equation}
where
\begin{eqnarray}
N_1 &=& \exp\left[-\frac{\mu^2}{2}(2l\delta+2l\epsilon+\delta^2+\epsilon^2) 
\right], \\
N_2 &=& \exp\left[-\mu^2\delta\epsilon \right], \\
D_1 &=& \exp\left[-\mu^2(2l\delta+\delta^2) \right], \\
D_2 &=& \exp\left[-\mu^2(2l\epsilon+\epsilon^2)\right].
\end{eqnarray}
The mass squared difference is
\begin{equation}
\Delta m^2 = 2\rho^2 e^{-\mu^2l^2}N_1\left(1+N_2\right)
\sqrt{1+\frac{1}{\tan^2 2\theta}}.
\end{equation}
Imposing large mixing\footnote{For concreteness we 
consider large mixing to be $|\tan 2\theta|\ge 2.3$, which agrees with the 99\% CL 
contours obtained by the neutrino oscillation analysis of the SuperKamiokande 
atmospheric data \cite{SK_atm}.}, one can safely write
\begin{equation}
\Delta m^2\simeq 2\rho^2 e^{-\mu^2l^2}N_1\left(1+N_2\right).
\label{mass1}
\end{equation}
Note that in this approximation $\Delta m^2$ is equal to the numerator
of $\tan 2\theta$ times $\rho^2e^{-\mu^2 l^2}$. 
This important observation is valid in all cases we will discuss
henceforth, as long as $\tan^2 2\theta\gg 1$ (which is the condition we would like
to meet). 

We start by obtaining bounds for $l$. Since the largest element of $m_{\nu}$ is 
the (22) element, imposing an upper bound on the heaviest neutrino mass of 
$m^2\lesssim 10$~eV$^2$ leads to $l> 7.08\mu^{-1}$. On the other hand, from 
Eq.~(\ref{mass1}) and noting that $N_1,N_2\le 1$, one obtains a rough upper bound 
for $l$ by requiring $\Delta m^2> 10^{-3}$~eV$^2$: $l\lesssim 7.8\mu^{-1}$. 
This can also be derived directly from Eq.~(\ref{case1}) by
noting that the largest neutrino mass matrix element should be (roughly) larger
than the lower bound on $\sqrt{\Delta m^2}$ imposed by the atmospheric neutrino 
data. It is important to comment that for values of $\mu l$ close to the upper
bound mentioned above, the neutrino masses are hierarchical, while for $\mu l$ much 
smaller than $7.7$, the eigenvalues of the neutrino mass matrix are degenerate and
much larger than $\sqrt{\Delta m^2}$. 

Next, we determine what values of $\epsilon$ and $\delta$, for fixed $l$, are required 
in order to obtain large mixing and the appropriate $\Delta m^2$ range. 
For fixed $l$, $\Delta m^2$ fixes the order of 
magnitude of $2N_1(1+N_2)$ (see Eq.~(\ref{mass1})), which is the numerator
of $\tan 2\theta$ (see Eq.~(\ref{tan1})). Large mixing requirements then determine
the order of magnitude of $(D_1-1)$ (note that $D_2+1$ is $O(1)$), which allows
one to compute an upper bound for $\delta$. Once this is done, one revisits the 
previous expressions and computes $\epsilon$. Table~\ref{table1} contains 
values of $\mu\epsilon$ and $(\delta/l)_{max}$ as a function of $\mu l$. 

\begin{table}[t]
\caption{{\it Case A} -- 
The allowed range of $\mu\epsilon$ and the maximum allowed value
for $\delta/l$, for different values of $\mu l$, which satisfy $1\times
10^{-3}~{\rm eV^2}<\Delta m^2<8\times 10^{-3}~{\rm eV^2}$ and $|\tan 2\theta|>2.3$ 
\cite{SK_atm}.}
\label{table1}
\begin{center}
\begin{tabular}{|c|c|c|} \hline
$\mu l$ & $\mu\epsilon$ & $(\delta/l)_{\rm max}$ \\ \hline \hline
7.1 & $\sim$ 1.1 -- 1.3 & $10^{-6}$ \\ \hline
7.2 & $\sim$ 0.9 -- 1.2 & $10^{-5}$ \\ \hline
7.3 & $\sim$ 0.7 -- 1.0 & $10^{-4}$ \\ \hline
7.4 & $\sim$ 0.5 -- 0.8 & $10^{-4}$ \\ \hline
7.5 & $\sim$ 0.3 -- 0.6 & $10^{-3}$ \\ \hline
7.6 & $\sim$ 0.1 -- 0.4 & $10^{-3}$ \\ \hline
7.7 & $\sim$ 0.0 -- 0.2 & $10^{-3}$ \\ \hline
\end{tabular}
\end{center}
\end{table} 

In order to examine how ``finely-tuned'' this scenario is, we look at the maximum
value for the ratio $\delta/l$, which is how much the distance $|l_{\mu}-n_{\mu}|$
is allowed to deviate from $|l_{\tau}-n_{\tau}|$, normalised by the smallest distance.
One can see that  $(\delta/l)_{\rm max} \lesssim10^{-3}$ over all possible values of 
$l,\epsilon$, for small $\epsilon$. It is curious that the fine-tuning gets more
severe as $\epsilon$ increases. This can be understood easily by the following line
of reasoning: $\Delta m^2$ essentially constrains $\epsilon$ due to the fact that,
it turns out, $\epsilon\gg \delta$; for small values of $l$, a large suppression
is required in order to obtain the correct value for $\Delta m^2$, which must
come from roughly $e^{-\mu^2l\epsilon}$. Hence bigger values of $\epsilon$
are required. However, larger values of $\epsilon$ suppress the numerator of
Eq.~(\ref{tan1}), requiring a larger suppression in the denominator, which is
mostly a function of $l$ and $\delta$, and the largest allowed value for $\delta$
decreases.

{\it Case B} -- Fig.~\ref{geoms}(B) leads to the Dirac mass matrix
\begin{equation}
m_{\nu}=\rho \left(\matrix{e^{-\frac{\mu^2}{2}(2l+\delta+\alpha)^2} & 
e^{-\frac{\mu^2}{2}l^2} 
\cr e^{-\frac{\mu^2}{2}\alpha^2} & e^{-\frac{\mu^2}{2}(l+\delta)^2}}\right),
\label{case2}
\end{equation} 
where $l_{21}\equiv l$ and $\l_{22}\equiv l+\delta$. 
We first assume $\delta>0$. This leads, to a very good approximation (for 
$\delta\ll l$), to
\begin{equation}
\tan 2\theta \simeq \frac{2\exp\left[-\frac{\mu^2}{2}(2l^2+2l\delta+\delta^2) 
\right]}
{\exp[-\mu^2l^2]-\exp[-\mu^2\alpha^2]-\exp[-\mu^2(l+\delta)^2]},
\label{tan2}
\end{equation} 
and, remembering that we will require $\tan^22\theta$ to be large,
\begin{equation}
\Delta m^2 \simeq 2\rho^2 \exp\left[-\frac{\mu^2}{2}(2l^2+2l\delta+\delta^2) \right].
\label{mass2}
\end{equation}

It is clear that the same approximate bounds on $l$, which applied in Case A, also
apply here, and again we analyse what constraints are imposed on the other parameters
for fixed values of $l$ within the allowed range. It should also be noted that,
similarly to Case A, the charged lepton mass matrix is safely diagonal, given that,
it turns out, $l_{21},l_{22}\gtrsim 7\mu^{-1}$.

For fixed $l$, the $\Delta m^2$ constraint fixes the range of $\delta$ and 
the numerator of $\tan 2\theta$ (Eq.~(\ref{tan2})). 
The smallness of this term will determine how much the terms
in the denominator of $\tan 2\theta$ are required to cancel in order to obtain 
large mixing. The only free parameter which remains is $\alpha$.
Table~\ref{table2} illustrates the severe fine-tuning between the value of 
$l$ and $\alpha$ which is required in order to obtain large mixing. Only for $l$
close to its upper bound is $\delta$ small enough to allow for a near perfect 
cancellation between the terms $e^{-\mu^2l^2}$ and $e^{-\mu^2(l+\delta)^2}$ in 
Eq.~(\ref{tan2}), in which case $\alpha$ plays no significant role, provided that 
$e^{-\mu^2\alpha^2}$ is small enough (which is satisfied for $\mu\alpha\gtrsim 7.8$).
In this case, therefore, there is no need to tune the values of $\alpha$ and 
$l_{22}$ (this happens for $\mu l=7.7$ and $\mu\delta\lesssim 0.05$). 
This does not mean, however, that there is no fine-tuning, since one is required
to tune the values of $l_{21}$ and $l_{22}$ (for $\mu l=7.7$, 
$(\delta/l)_{\rm max}\lesssim 10^{-3}$).

\begin{table}[t]
\caption{{\it Case B} -- 
The allowed range of $\mu\delta$ and the maximum allowed value
for $|\alpha-l|/l$, for different values of $\mu l$, which satisfy $1\times
10^{-3}~{\rm eV^2}<\Delta m^2<8\times 10^{-3}~{\rm eV^2}$ and $|\tan 2\theta|>2.3$
\cite{SK_atm}.}
\label{table2}
\begin{center}
\begin{tabular}{|c|c|c|} \hline
$\mu l$ & $\mu\delta$ & $(|\alpha-l|/l)_{\rm max}$ \\ \hline \hline
7.1 & $\sim$ 1.0 -- 1.3 & $10^{-6}$ \\ \hline
7.2 & $\sim$0.8 -- 1.1 & $10^{-5}$ \\ \hline
7.3 & $\sim$0.6 -- 0.9 & $10^{-5}$ \\ \hline
7.4 & $\sim$0.4 -- 0.7 & $10^{-4}$ \\ \hline
7.5 & $\sim$0.2 -- 0.5 & $10^{-3}$ \\ \hline
7.6 & $\sim$0.05 -- 0.3 & $10^{-2}$ \\ \hline
7.7 & $\sim$0.5 -- 0.1 & $10^{-2}$ \\ 
 & $\sim$0 -- 0.05 & unconstrained \\ \hline
\end{tabular}
\end{center}
\end{table} 

Next, we examine the case $l_{21}>l_{22}$, {\it i.e.}\/ $\delta<0$. 
In this case, the dominant term in the denominator of Eq.~(\ref{tan2})
is $e^{-\mu^2(l-|\delta|)^2}$, which cannot be cancelled by
appropriately choosing $\alpha$. The only way to obtain large mixing is, therefore,
to choose $|\delta|\ll l$ and $\alpha$ large. Numerically, we obtain that
$\mu|\delta|$ cannot be larger than 0.05. The bounds on both $\Delta m^2$ and
$\tan2\theta$ can only be simultaneously satisfied for $\mu l_{22}=7.6$,
$0.04\lesssim \mu|\delta| \lesssim 0.05$ and $\mu l_{22}=7.7$ and 
$0\lesssim \mu|\delta| \lesssim 0.05$, while $\mu\alpha\gtrsim 7.8$. 
In this case, the fine-tuning occurs between $l_{21}$ and $l_{22}$, which 
have to be equal to one another to more than one part in 10$^{2}$. 

{\it Case C} -- Fig.~\ref{geoms}(C) leads to the Dirac mass matrix
\begin{equation}
m_{\nu}=\rho \left(\matrix{e^{-\frac{\mu^2}{2}(l+\delta+\epsilon)^2} & 
e^{-\frac{\mu^2}{2}l^2} 
\cr e^{-\frac{\mu^2}{2}(l+\delta)^2} & e^{-\frac{\mu^2}{2}(l+\epsilon)^2}}\right),
\label{case3}
\end{equation}  
which yields the approximate relations ($\epsilon,\delta\ll l$)
\begin{equation}
\tan 2\theta \simeq \frac{2N'_{1}[1+N'_{2}]}
{(1-D'_{1})(1-D'_{2})},
\label{tan3}
\end{equation}
and
\begin{equation}
\Delta m^2 \simeq 2\rho^2 e^{-\mu^2l^2}N'_{1}[1+N'_{2}],
\label{mass3}
\end{equation}
where 
\begin{eqnarray}
N'_1 &=& \exp\left[-\frac{\mu^2}{2}(2l\epsilon+\epsilon^2) 
\right], \\
N'_2 &=& \exp\left[-\mu^2(2l\delta + \delta^2 + \delta\epsilon) \right], \\
D'_1 &=& \exp\left[-\mu^2(2l\delta+\delta^2) \right], \\
D'_2 &=& \exp\left[-\mu^2(2l\epsilon+\epsilon^2)\right].
\end{eqnarray}
Note that the terms in the square bracket in Eqs.~(\ref{tan3},\ref{mass3}) are,
by definition, between 1 and 2 and play no significant role in order of magnitude
discussions. As in the two previous cases, $l$ is bound between roughly $7.1\mu^{-1}$
and $7.8\mu^{-1}$ (assuming the ``atmospheric'' $\Delta m^2$ and that 
$m_{\nu}^2<10$~eV$^2$) and again we discuss the required cancellation among 
parameters
for fixed values of $l$. As before, the requirement that
$\Delta m^2$ satisfies the atmospheric neutrino data determines the order of
magnitude of the numerator of $\tan 2\theta$ and, in this case, also determines
the value of $\epsilon$. The requirement $|\tan 2\theta|>2.3$ is met by choosing
$(1-D'_{1})$ very small, which implies that $\delta$ has to be very small. 
Table~\ref{table3} 
contains the value of $\mu l$, $\mu\epsilon$, and ($\delta/l)_{\rm max}$
which are obtained when trying to fit large mixing in the 2-3 sector with 
$\Delta m^2\sim 10^{-3}$~eV$^2$. 

\begin{table}[t]
\caption{{\it Case C} -- 
The allowed range of $\mu\epsilon$ and the maximum allowed value
for $\delta/l$, for different values of $\mu l$, which satisfy $1\times 
10^{-3}~{\rm eV^2} <\Delta m^2<8\times 10^{-3}~{\rm eV^2}$ and $|\tan 2\theta|>2.3$
\cite{SK_atm}.}
\label{table3}
\begin{center}
\begin{tabular}{|c|c|c|} \hline
$\mu l$ & $\mu\epsilon$ & $(\delta/l)_{\rm max}$ \\ \hline \hline
7.1 & $\sim$ 1.1 -- 1.3 & $10^{-6}$ \\ \hline
7.2 & $\sim$ 0.9 -- 1.2 & $10^{-5}$ \\ \hline
7.3 & $\sim$ 0.7 -- 1.0 & $10^{-4}$ \\ \hline
7.4 & $\sim$ 0.5 -- 0.8 & $10^{-4}$ \\ \hline
7.5 & $\sim$ 0.3 -- 0.6 & $10^{-3}$ \\ \hline
7.6 & $\sim$ 0.1 -- 0.4 & $10^{-3}$ \\ \hline
7.7 & $\sim$ 0.05 -- 0.2  & $10^{-1}$ \\ 
 & $\sim$ 0 -- 0.05 & unconstrained \\ \hline
\end{tabular}
\end{center}
\end{table} 

It is noteworthy that in the limit $\epsilon\rightarrow 0$ there is no
constraint on $\delta$. In this limit, the two lepton doublets are on top 
of each other, and large mixing is guaranteed, regardless of the location
of the singlet fields. This situation, however, is in sharp contradiction with
the bound from 
the flavour changing $\tau\rightarrow\mu\gamma$ decay, as discussed earlier in
this section. Indeed, when we fit for the charged lepton masses imposing that
$\tau\rightarrow\mu\gamma$ is small, we find that $\mu\epsilon\gtrsim 1$, and
large mixing with the appropriate mass squared is only obtained for 
$\mu l\lesssim 7.3$ (and $(\delta/l)_{\rm max}\lesssim 10^{-5}$!).   
 
{\it Case D} -- Fig.~\ref{geoms}(D) leads to the Dirac mass matrix
\begin{equation}
m_{\nu}=\rho \left(\matrix{e^{-\frac{\mu^2}{2}(l+\alpha+\beta)^2} & 
e^{-\frac{\mu^2}{2}(l+\alpha)^2} 
\cr e^{-\frac{\mu^2}{2}(l+\beta)^2} & e^{-\frac{\mu^2}{2}l^2}}\right),
\label{case4}
\end{equation}  
which leads to the relations
\begin{equation}
\tan 2\theta = \frac{2N''_{1}[1+N''_{2}]}
{(D''_1-1)(D''_2+1)-(1-e^{-2\mu^2\alpha\beta})D''_1D''_2},
\label{tan4}
\end{equation}
and (for $\tan^2 2\theta$ large)
\begin{equation}
\Delta m^2 \simeq 2\rho^2 e^{-\mu^2l^2}N''_{1}[1+N''_{2}],
\label{mass4} 
\end{equation}
where 
\begin{eqnarray}
N''_1 &=& \exp\left[-\frac{\mu^2}{2}(2l\alpha+\alpha^2) 
\right], \\
N''_2 &=& \exp\left[-\mu^2(2l\beta + \beta^2 + \alpha\beta) \right], \\
D''_1 &=& \exp\left[-\mu^2(2l\alpha+\alpha^2) \right], \\
D''_2 &=& \exp\left[-\mu^2(2l\beta+\beta^2)\right].
\end{eqnarray}
As in the previous cases, $\mu l$ is naively bound between $\sim [7.1,7.8]$ simply
by requiring the largest neutrino mass squared to be less than 10~eV$^2$ and 
$\Delta m^2>10^{-3}$~eV$^2$. 
It is again useful to constrain the other parameters for fixed 
value of $\mu l$. In order to obtain a
correct value for $\Delta m^2$, each value of $l$ requires a specific range of
$\alpha$ (which determines $N''_1$, see Eq.~(\ref{mass4})). Once $\alpha$ is
fixed, we try to choose $\beta$ such that the mixing is large. In this case,
however, it is easy to see that changing $\beta$ for fixed $\alpha$ and $l$
does very little to increase $\tan 2\theta$, as it mainly affects the value 
of $D''_2$, which plays a diminished role when it comes to determining the
order of magnitude of $\tan 2\theta$ (see Eq.~(\ref{tan4})).

The only possibility for obtaining large mixing is to choose 
$\alpha$ such that $(D''_1-1)$
is very small, {\it i.e}\/ to choose $\alpha$ very small. Note that this choice
automatically renders the term in the denominator of Eq.~(\ref{tan4}) proportional
to $1-e^{2\mu^2\alpha\beta}$ small.\footnote{One may worry what happens if 
$\beta$ is very large. In this case $D''_2$ is tiny, which efficiently suppresses
this term.} Small values of $\alpha$ require $l$ to be large enough so that 
an appropriate value of $\Delta m^2$ is obtained. As a result, only values
$\mu l\gtrsim 7.6, \mu\alpha \lesssim 0.05$ potentially yield phenomenologically
allowed values for $\Delta m^2$ and $\tan 2\theta$.

As in Case C, the charged lepton sector imposes (here very severe)
constraints over the allowed values for the $|l_{\mu}-l_{\tau}|$ distance,
$\alpha$. As discussed in the beginning of this subsection, 
$\mu\alpha\gtrsim 1$ is required by $\tau\rightarrow\mu\gamma$ constraints,
which implies that there are {\sl no solutions} with 
$10^{-3}~{\rm eV^2}<\Delta m^2<8\times 10^{-3}$~eV$^2$, 
$(m_{\nu}^2)_{\rm max}<10$~eV$^2$, and $|\tan 2\theta|\ge 2.3$.    

In summary, we have analysed in detail the issue of obtaining large mixing and
$10^{-3}~{\rm eV^2}<\Delta m^2<8\times
10^{-3}$~eV$^2$ in the $\mu-\tau$ sector. For all
configurations we discussed, this is only achieved by carefully choosing the value
of specific distances. Quantitatively, we find that it is necessary that the
value of two {\it a priori}\/ independent distances be equal to, at least, 1 part
in 10$^{2}$ in order to obtain large mixing in the atmospheric neutrino 
sector. Cases where maximum mixing is ``naturally'' obtained
by placing fields ``on top'' of one another (Cases C and D), were ruled out by  
flavour changing neutral current constraints combined with obtaining the
observed values for the corresponding charged lepton masses.

We conclude with a comment on the order one higher dimensional Yukawa
couplings which we set to unity throughout. Firstly, note that these coefficients
can be readily ``absorbed'' by slightly modifying the distance between the
lepton fields. Explicitly,
\begin{eqnarray}
\alpha e^{-\frac{\mu^2}{2}l^2}&=&e^{-\frac{\mu^2}{2}(l')^2}, \\
2\ln{\alpha}-(\mu l)^2&=&-(\mu l')^2, \\
\frac{l}{l'}&=&\sqrt{1+\frac{2\ln{\alpha}}{(\mu l')^2}}, 
\end{eqnarray}
where $l/l'$ is how much the distance has to be changed in order to ``absorb'' the
coefficient $\alpha$. For neutrino masses, $(\mu l')^2\gtrsim 50$, such that, even if
$\alpha$ were as small as 0.1, 
$|l/l'-1|\lesssim 0.05$. Secondly, and most importantly, the presence of
the order one coefficients will not affect our fine-tuning discussions -- 
parameters just have to be slightly ``redefined.'' 
For example, in Case A, instead of having to ``tune'' $l_1$
to $l_2$, one would tune $l_1$ to $l_2$ plus a small, constant, coefficient
dependent on the ratio of the higher dimensional Yukawa couplings. 
The ``amount'' of fine-tuning remains the same.   

\subsection{Three Neutrino Families}

In this subsection, we discuss the case of three lepton families, concentrating on 
obtaining appropriate neutrino mass-squared differences and mixing angles for 
solving 
the atmospheric and solar neutrino puzzles, while satisfying the constraints 
from the CHOOZ and Palo Verde experiments \cite{CHOOZ}.\footnote{In this paper,
we will not try to accommodate the LSND data \cite{LSND}, 
which is still to be confirmed by another experiment, and will only consider 
three active neutrino species.}

Unlike the two-family case, in the case of three families it is not simple to
perform a search for all configurations which potentially yield the appropriate 
values for the oscillation parameters. Furthermore, after some configurations are 
identified, a general analytic discussion of ``fine-tuning'' among different 
parameters is not illuminating. For these reasons, we decided to approach the situation 
in a different way.

Initially, we numerically scanned the ``location space'' of the left-handed lepton 
doublets and the right-handed neutrino singlets, restricting ourselves to branes 
which are fat in only one extra dimension. During the scan, we looked for solutions
which yield neutrino masses and mixing angles such that all the neutrino 
experimental data could be accounted for by neutrino oscillations \cite{neutrinos}.
Details of the numerical scan are described in the Appendix, and some results
are the following: {\it i-}\/ As expected, obtaining reasonable values for the 
mixing angles is quite hard, more so than the correct order of magnitude for the
mass-squared differences. {\it ii-}\/ 
In all of the acceptable configurations, the $\mu$--$\tau$ system
falls into one of the configurations (see Fig.~\ref{geoms}) 
discussed in detail in the previous subsection, with the same fine-tuning characteristics. 
{\it iii-}\/ Only configurations which yield hierarchical neutrino masses 
were found.
{\it iv-}\/ For a significant fraction of the points
($\sim$50\%) flavour changing effects in the charged lepton sector are absent, 
and the charged lepton mass matrix is virtually diagonal. 
In summary, it was possible to find several
configurations which yield appropriate neutrino masses and mixing angles for 
explaining all neutrino data. 

\begin{figure}
\centerline{
    \psfig{file=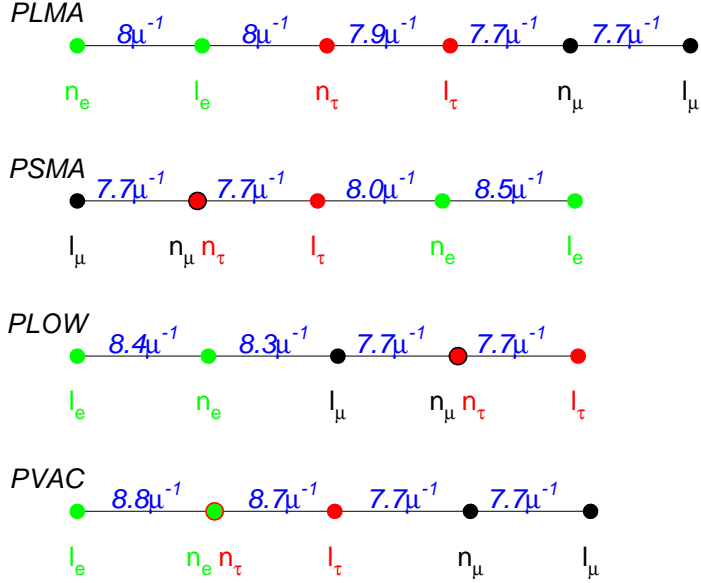,width=0.7\columnwidth}}
    \caption{Four examples which yield neutrino masses and mixing
angles which satisfy all neutrino data.}
    \label{4pontos}
\end{figure}

Four different configurations are depicted in 
Fig.~\ref{4pontos}.\footnote{For all four examples, the brane seems to be too fat.
This is a generic feature of almost all solutions we find, as discussed in the
Appendix, and should be interpreted as a motivation for introducing more than one
``fat'' extra-dimension.}
These are labelled PLMA, PSMA, PLOW and PVAC, alluding
to the fact that they will fall into the LMA, SMA, LOW and VAC solutions 
\cite{neutrinos} 
to the solar neutrino puzzle, respectively. Table~\ref{pontos} contains
the exact distances for the four configurations.\footnote{see the 
Appendix for a definition of the ``distance parameters.''}  
Note that PLMA and PVAC have the
second and third generation configuration depicted in Fig.~\ref{geoms}(B), 
while PSMA and PLOW present that depicted in Fig.~\ref{geoms}(A). 

\begin{table}[t]
\caption{Points which yield neutrino masses and mixing angles such that all
neutrino data can be accommodated. The different points fall (PLMA, PSMA, PLOW,
PVAC) in different solutions to the solar neutrino puzzle, and are labelled 
accordingly. For the definition of the parameters $l_1,l_2,l_3,x,y$, see
Fig.~\ref{3gen_draw} in the Appendix.}
\label{pontos}
\begin{center}
\begin{tabular}{|c|c|c|c|c|c|} \hline
configuration & $\mu l_1$ & $\mu l_2$ & $\mu l_3$ & $\mu x$ & $\mu y$ \\ \hline \hline
PLMA & 8.0 & -7.7 & 7.9 & -16.0 & -15.6 \\ \hline
PSMA & 8.5 & 7.7 & 7.7 & 15.7 & 0.0 \\ \hline
PLOW & -8.4 & 7.7 & 7.7 & -16.0 & 0.0 \\ \hline
PVAC & -8.8 & -7.7 & 8.7 & 0.0 & -16.4 \\ \hline
\end{tabular}
\end{center}
\end{table}
In order to try to quantify if/how these solutions are finely-tuned, we study the
stability of the four configurations depicted in Fig.~\ref{4pontos} by
probing ``around'' them. Explicitly, we fix the ``smallest'' parameter (
$l_2$ in the case of PSMA and PLOW and
$l_3$ in the case of PSMA and PLOW) and vary the other ones ($l_1,x,y,$ and 
$l_2$ or $l_3$) from $p_i-0.1$ to $p_i+0.1$ in steps of 0.02, where
$p_i$ is the original value of each of the four parameters (for a total of 11$^4$
points). The size of the interval is guided
by the results of our two-generation studies in the previous subsection, {\it i.e.},
we choose an interval which is small enough so that the large atmospheric 
angle is not disturbed too much.

\begin{figure}
\centerline{
    \psfig{file=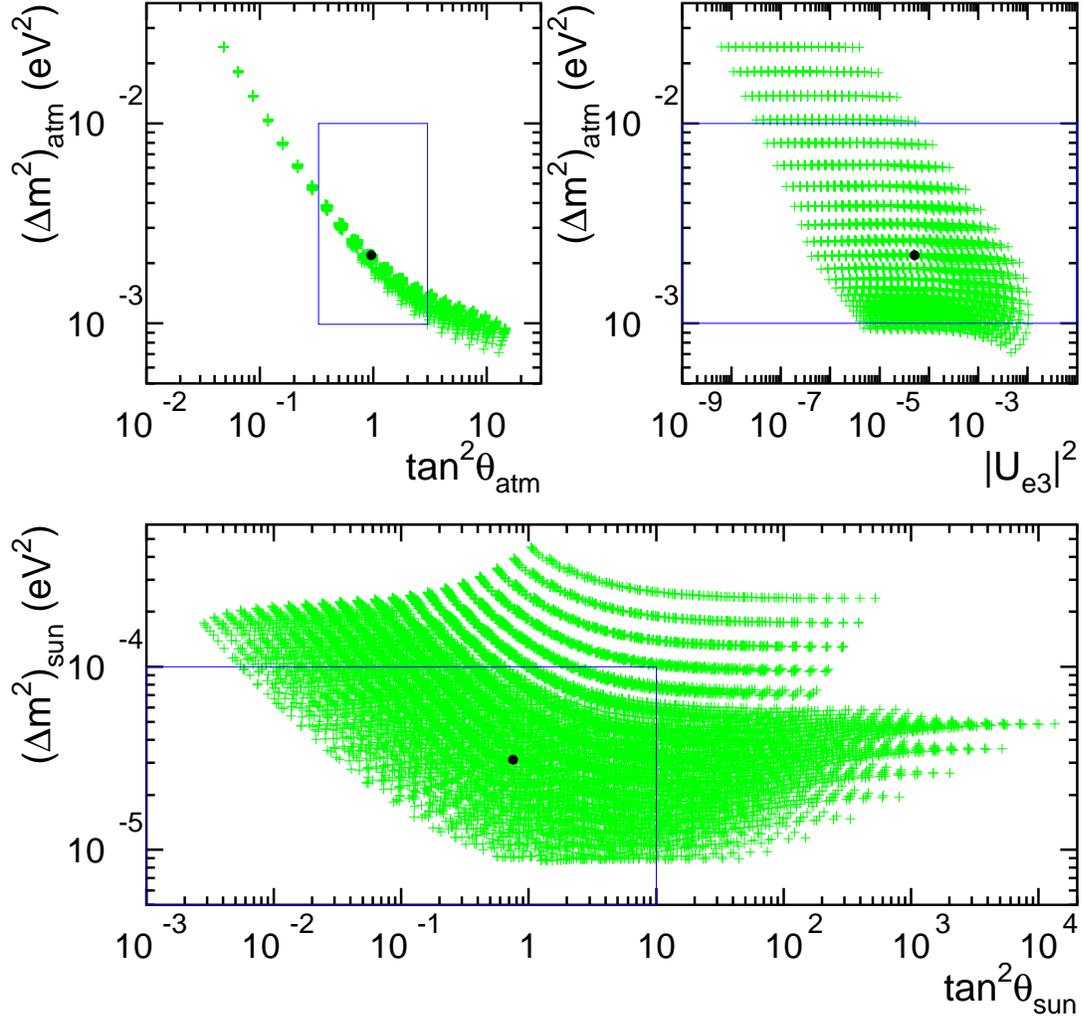,width=1.0\columnwidth}}
    \caption{Values of $\Delta m^2_{\rm atm} \times \tan^2\theta_{\rm atm}$ 
(top, left), $\Delta m^2_{\rm atm} \times |U_{e3}|^2$ (top right), and
$\Delta m^2_{\odot} \times \tan^2\theta_{\odot}$ (bottom) which are obtained
when perturbing around PLMA (see text for details). The results obtained at PLMA
is indicated by a solid, dark dot.}
    \label{ponto_lma}
\end{figure}

\begin{figure}
\centerline{
    \psfig{file=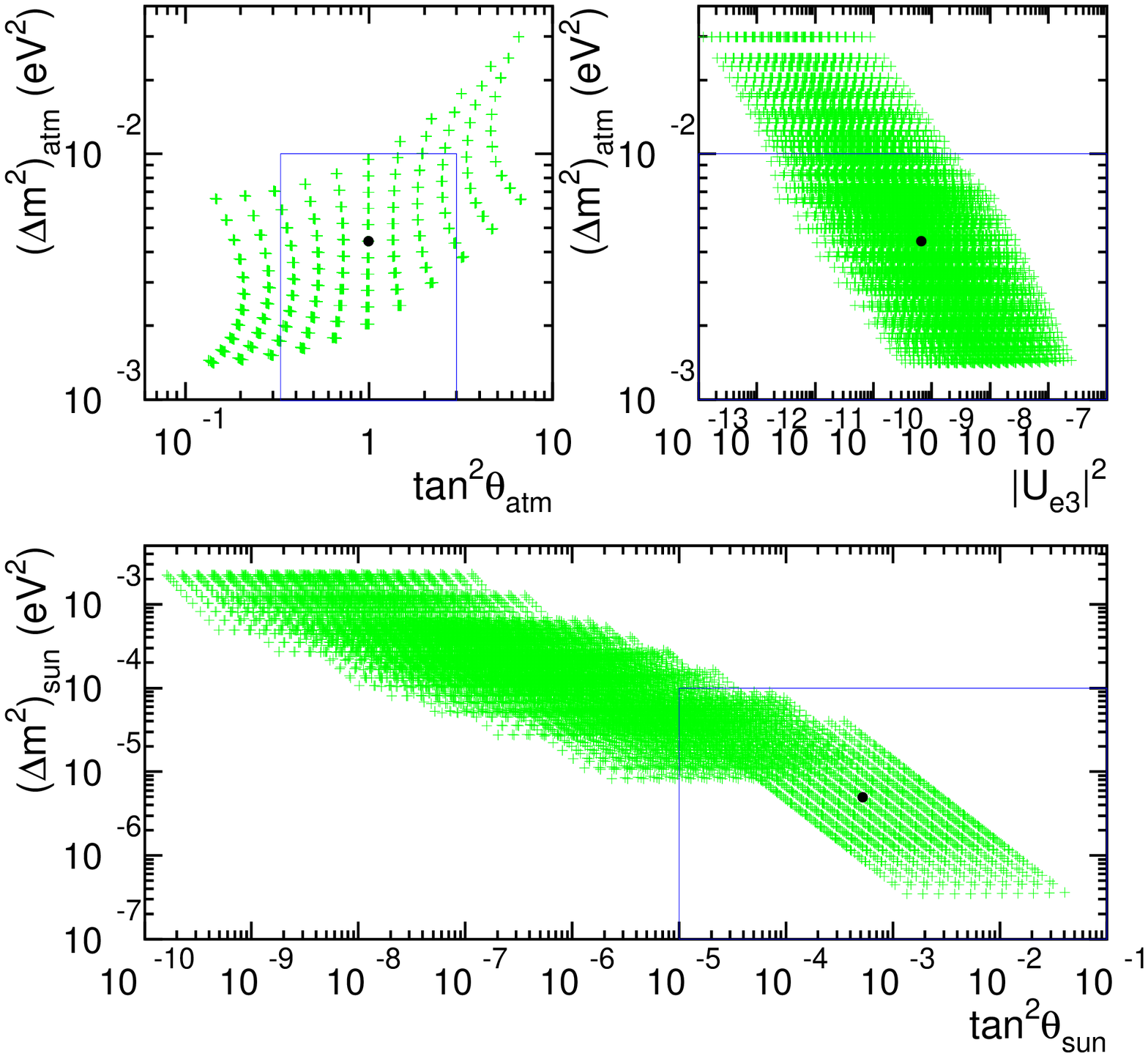,width=1.0\columnwidth}}
    \caption{Same as Fig.~\ref{ponto_lma}, when perturbing around PSMA.}
    \label{ponto_sma}
\end{figure}

Fig.~\ref{ponto_lma} (Fig.~\ref{ponto_sma}) depicts the result of this stability
study around PLMA (PSMA). 
In the figures, we plot the values obtained for all the oscillation parameters:
$\Delta m^2_{\rm atm} \times \tan^2\theta_{\rm atm}$ (top left), 
$\Delta m^2_{\rm atm} \times |U_{e3}|^2$ (top right), and
$\Delta m^2_{\odot} \times \tan^2\theta_{\odot}$ (bottom). The thin [blue] boxes 
mark the ``allowed regions,'' as defined in the Appendix, while the
solid, dark dot indicates the results obtained for PLMA (PSMA). When perturbing
about PLOW and PVAC, similar results are obtained (indeed, we have repeated
this procedure for a large number of points).
 
The results contained in the figure are quite interesting. First, it should be
noted that, as expected, the atmospheric parameters do  not vary substantially. 
Numerically, 56\% (54\%) of the points remain inside the ``atmospheric box'', 
$10^{-3}~{\rm eV^2}<\Delta m^2_{\rm atm}<10^{-2}~{\rm eV^2}$, $1/3<\tan^2
\theta_{\rm atm}<3$ in the case of PLMA (PSMA). On the other hand, some of the
solar parameters and $|U_{e3}|^2$ vary drastically. For example, while perturbing 
around PLMA (PSMA), $\tan^2\theta_{\odot}$ varies by 7 (9) orders of magnitude, and
similar results apply to $|U_{e3}|^2$. Moreover, $\Delta m^2_{\odot}$, while somewhat 
stable around PLMA (it varies by 2 orders of magnitude), spans 5 orders of magnitude around
PSMA.  

Does this observed numerical instability indicate a fine-tuning among different
parameters? After all, what we would like to learn from these 
``stability studies'' 
is whether, in order to guarantee that
the solar parameters and $|U_{e3}|^2$ are inside a certain range, the distances
between some lepton fields have to be chosen very carefully, and whether slight 
modifications
severely spoil the result. In order to verify that this is the case, we look at 
PLMA and PSMA in more detail. The PLMA-like configurations 
(see Fig.~\ref{4pontos}) lead to the following neutrino Dirac mass matrix:
\begin{equation}
m_{\nu}=\left(\matrix{a & 0 & a'
\cr 0 & b & 0 \cr 0 & b' & c}\right),
\label{dirac_LMA}
\end{equation}  
where $a,a',b,b',c$ are related, respectively, to the distances $|l_e-n_e|$,
$|l_e-n_{\tau}|$,
$|l_{\mu}-n_{\mu}|$, $|l_{\tau}-n_{\mu}|$, $|l_{\tau}-n_{\tau}|$. 
Eq.~(\ref{dirac_LMA}) 
leads to
\begin{equation}
m_{\nu}m_{\nu}^{\dagger}=\left(\matrix{a^2+a'^2 & 0 & a'c
\cr 0 & b^2 & bb' \cr a'c & bb' & b'^2+c^2}\right).
\label{m2_LMA}
\end{equation}  
From the discussion in the previous subsection, we know that large mixing in the 
atmospheric sector is 
obtained for $b\simeq b'> c$ (Case B). It is also easy to see that, in order to
obtain $\Delta m^2_{\odot}<\Delta m^2_{\rm atm}$, $a,a',c<b,b'$. In light of this, we
multiply Eq.~(\ref{m2_LMA}) by a $\theta$ rotation about the ``$e$-axis'' such that
$\sin{\theta}\equiv b'/\sqrt{b^2+b'^2}$
and obtain
\begin{eqnarray}
& \left(\matrix{1 & 0 & 0
\cr 0 & \cos\theta & \sin\theta \cr 0 & -\sin\theta & 
\cos\theta} \right)
\left(\matrix{a^2+a'^2 & 0 & a'c
\cr 0 & b^2 & bb' \cr a'c & bb' & b'^2+c^2}\right)
\left(\matrix{1 & 0 & 0
\cr 0 & \cos\theta & -\sin\theta \cr 0 & \sin\theta & 
\cos\theta} \right)\simeq \nonumber \\
& \left(\matrix{a^2+a'^2 & a'c\sin\theta & a'c\cos\theta
\cr a'c\sin\theta & b^2+b'^2+c^2\sin^2\theta & c^2\sin\theta\cos\theta \cr 
a'c\cos\theta & 
c^2\sin\theta\cos\theta & c^2\cos^2\theta }\right).
\end{eqnarray} 
If $b^2,b'^2\gg c^2,a'c$, it is easy to see that $\theta\simeq\theta_{\rm atm}$ 
(corrections to this lead to a nonzero $U_{e3}$, for example), as expected. Furthermore,
the two other eigenstates are ``light'' (as long as $a,a',c$ are ``small'') 
and the solar angle is easily calculable:
\begin{equation}
\cos 2\theta_{\odot}\simeq \frac{c^2\cos^2\theta-(a^2+a'^2)}
{\sqrt{(a^2+a'^2-c^2\cos^2\theta)^2+4a'^2c^2\cos^2\theta}}.
\end{equation}
In order to obtain, for example, large mixing in the solar sector, it is necessary 
to satisfy $a^2+a'^2\simeq c^2\cos^2\theta \simeq c^2/2$. This is what 
happens at PLMA: $a=a'=e^{-0.5\times 8.0^2}$ and $c=e^{-0.5\times 7.9^2}$, such that 
$a/c\simeq 0.45$. Therefore, 
a large solar angle is a consequence of finely tuning the distances
$|l_{e}-n_{e,\tau}|$ and $|l_{\tau}-n_{\tau}|$. It should be noted that the results of 
the scan involve further complications, such as $c\sim b'$, which leads to  
large values of $\Delta m^2_{\odot}$ and significant 
corrections to the atmospheric angle. 

The PSMA-like configurations (see Fig.~\ref{4pontos}) lead to
the following neutrino Dirac mass matrix:
\begin{equation}
m_{\nu}=\left(\matrix{\epsilon & 0 & 0
\cr 0 & b & b' \cr a & b'' & b'''}\right),
\label{dirac_SMA}
\end{equation}  
where $\epsilon,a, b,b',b'',b'''$ are related, respectively, to the distances 
$|l_e-n_e|$, $|l_{\tau}-n_{e}|$,
$|l_{\mu}-n_{\mu}|$, $|l_{\mu}-n_{\tau}|$, $|l_{\tau}-n_{\mu}|$, $|l_{\tau}-n_{\tau}|$. 
From Fig.~\ref{4pontos}, it is easy to see that $\epsilon\ll a \ll b,b',b'',b'''$. 
Eq.~(\ref{dirac_SMA}) leads to
\begin{equation}
m_{\nu}m_{\nu}^{\dagger}\simeq\left(\matrix{\epsilon^2 & 0 & a\epsilon
\cr 0 & b^2+b'^2 & bb''+b'b''' \cr a\epsilon & bb''+b'b''' & b''^2+b'''^2 + a^2} 
\right).
\label{m2_SMA}
\end{equation}    
As before, the atmospheric angle and mass difference result
from diagonalizing the $\mu$--$\tau$ sector. As we did in the PLMA case 
above, we can ``undo'' the atmospheric rotation and obtain
\begin{equation}
U_{\theta} m_{\nu}m_{\nu}^{\dagger}U^{\dagger}_{\theta}= 
 \left(\matrix{\epsilon^2 & a\epsilon\sin\theta & a\epsilon\cos\theta
\cr a\epsilon\sin\theta & m_2^2+a^2\sin^2\theta & a^2\sin\theta\cos\theta \cr 
a\epsilon \cos\theta & a^2\sin\theta\cos\theta & m_1^2+a^2\cos^2\theta }\right),
\end{equation} 
where $m_1^2,m_2^2$ are the lightest and heaviest eigenstates obtained after 
diagonalizing the $\mu$--$\tau$ sector. They are given by
\begin{equation}
m_{1,2}^2= \frac{b^2+ b'^2+b''^2+b'''^2}{2}\pm \frac{1}{2}\sqrt{(b^2+ b'^2-b''^2-
b'''^2)^2+4(bb''+b'b''')^2}.
\end{equation}
Large mixing in the atmospheric sector requires $b\sim b' \sim b'' \sim b'''$ 
(see Case A in the previous subsection), and, in the
case of hierarchical neutrinos (the case of interest here), $\Delta m^2_{\rm atm}
\sim m_2^2 > m_1^2$. In this case, it is trivial to compute the solar angle and 
mass squared difference, keeping in mind that $a^2\gg \epsilon^2$ and that 
$\cos^2\theta_{\rm atm}\simeq 1/2$:
\begin{eqnarray}
\Delta m^2_{\odot}&\simeq & a^2/2+m_1^2, \\
\tan\theta_{\odot} \simeq \sin\theta_{\odot} &\simeq& \frac{a\epsilon/\sqrt{2}}
{a^2/2+m_1^2}.
\end{eqnarray} 
Exactly at PSMA, $m_1^2=0$, and $\Delta m^2_{\odot}$ and $\tan^2\theta_{\odot}$ are
governed by $a$ and $\epsilon$. As long as this is the case ($m^2_1\ll a^2$), varying
$l_1,l_3$ and $x$ (which changes the distances $|l_e-n_{e}|$ and $|l_{\tau}-n_{e}|$) only
perturbs around PLMA. However, when $m_1^2$ grows (this happens for 
$y\neq 0$) the situation changes dramatically, and $\Delta m^2_{\odot}$ grows, while 
$\tan^2\theta_{\odot}$ decreases sharply. This explains the ``cloud'' of points 
in Fig.~\ref{ponto_sma}(bottom) at $10^{-5}~{\rm eV^2}\lesssim\Delta m^2_{\odot}
\lesssim 3\times 10^{-3}~{\rm eV^2}$, $10^{-10}\lesssim \tan^2\theta_{\odot}\lesssim
10^{-3}$. In summary, the solar parameters depend very strongly on $y$, the distance
between $n_{\mu}$ and $n_{\tau}$. More so than the atmospheric parameters!
Note that a very similar behaviour is observed when perturbing around PLOW, where
the $\mu$--$\tau$ fields also fall into the configuration discussed in Case A in the
previous subsection.

We conclude this subsection with a summary of the results and a word of caution.
We have numerically scanned the one-dimensional parameter space of three neutrino
generations and encountered, out of many ``good'' candidates (which yield
small neutrino masses), only a handful of points which could potentially 
accommodate the neutrino data. 
As expected from our qualitative discussion at the 
beginning of this section, the great majority of the points ($>99.9\%$) fails to 
yield large enough mixing in the solar and/or atmospheric sectors. The situation 
is particularly constrained after $|U_{e3}|^2$ is required to be small. 
All the points identified had configurations and fine-tuning characteristics
similar to the ones discussed in the previous subsection (see Fig.~\ref{geoms}), 
with different choices for the position of the first generation fields. 

We selected four examples consistent with the different solutions to the solar neutrino
puzzle, and studied the stability of these solutions in order to address how 
``special'' the distances between lepton fields have to be in order to obtain 
a particular result. We find that small modifications (of order a few percent of 
the typical distance scale) seem to maintain
the large mixing in the atmospheric sector, in agreement with the discussion in the
previous subsection, while changing the solar parameters and $|U_{e3}|^2$ by 
many orders of magnitude. For two specific cases we identified the origin of this
effect. Therefore, it seems that obtaining appropriate solar parameters requires
tuning more parameters (or the same parameters more tightly) than what is already
required in order to obtain large mixing in the atmospheric sector.

It should be emphasised that we have not attempted a detailed study of the three by
three neutrino mass matrix in order to study correlations between the various 
distances, but decided to base our conclusions on a few ``case studies.''
It is not clear to us whether such a study is practical and/or illuminating.
Therefore, the conclusions we reached, although quite plausible and reasonable, 
should be read with some caution. 
   
\setcounter{equation}{0}
\setcounter{footnote}{0}
\section{Neutrino Masses without Right-Handed Neutrinos}

As mentioned in Sec.~2, the traditional see-saw mechanism is not an option for
generating very small neutrino masses in the 
case of models with a small quantum gravity scale. Nonetheless, one may still
obtain small {\sl Majorana} neutrino masses in the ADD scenario. The idea,
proposed in \cite{no_brane}, is to take advantage of the ``infrared desert'' 
\cite{ir_desert} to obtain very small symmetry breaking effects. More explicitly,
the idea is to impose lepton number as a conserved global symmetry, which is 
broken at some far away brane. Lepton number violating effective operators
would be generated in our brane, suppressed by $e^{-mr}$ where $m$ is some
typical mass scale and $r$ is the distance between the branes (see \cite{no_brane,
ir_desert} for details).

For our purposes, it is enough to assume that the following Majorana neutrino
mass matrix is generated after electroweak symmetry breaking:
\begin{equation}
M_{\nu}=m_{\rm Maj}
\left(\matrix{\alpha_{ee} & \alpha_{e\mu}e^{-\frac{\mu^2}{2}(l_{e}-l_{\mu})^2}
 & \alpha_{e\tau}e^{-\frac{\mu^2}{2}(l_e-l_{\tau})^2} \cr 
\alpha_{e\mu}e^{-\frac{\mu^2}{2}(l_{e}-l_{\mu})^2}  & \alpha_{\mu\mu} & 
\alpha_{\mu\tau}e^{-\frac{\mu^2}{2}(l_{\mu}-l_{\tau})^2} \cr 
\alpha_{e\tau}e^{-\frac{\mu^2}{2}(l_e-l_{\tau})^2}  & 
\alpha_{\mu\tau}e^{-\frac{\mu^2}{2}(l_{\mu}-l_{\tau})^2}
& \alpha_{\tau\tau}}\right),
\end{equation}    
where $m_{\rm Maj}$ is the overall Majorana mass scale, $\alpha_{\beta\gamma}$ 
are the higher dimensional order one couplings.\footnote{We are assuming, according 
to the general philosophy of the rest of the paper, that no flavour structure comes
from the lepton number violating sector. Such a structure (conserved
flavour symmetries, for example) would manifest itself by imposing relations between the 
$\alpha_{\beta\gamma}$.} The distances $|l_i-l_j|$ are the same
distances between the different left-handed lepton fields discussed in the
previous sections. 

The issue we would like to deal with in this section is whether
phenomenologically acceptable Majorana mass matrices can be obtained,  
if one assumes that the absence of flavour violating decays of the muon and 
the tau and the hierarchy of the charged lepton masses are explained by 
the AS scenario.

In the absence of the flavour changing constraints, it is clear that solutions
can be found: if one places all lepton doublets on top of one another, large
mixing in the solar and atmospheric sector arises quite naturally, and even a
small $|U_{e3}|^2$ and moderate hierarchy in the masses-squared can be obtained 
\cite{anarchy}.\footnote{Indeed, in this case large mixing would also come
from the charged lepton sector.} With the flavour changing constraints, however,
the situation is more involved, since the lepton doublet fields {\sl cannot}
sit on top of one another, as we argued in Sec.~3. Note that if all
distances were $|l_{\alpha}-l_{\beta}|\sim 2\mu^{-1}$ all off-diagonal terms
would be one order of magnitude smaller than the diagonal terms,\footnote{According
to the authors of \cite{NuSh}, gauge interactions may significantly alter
this picture. For simplicity, however, we will ignore these effects, which 
seem to be very dependent on the fermion localising mechanism.} and
obtaining large mixing in the atmospheric sector, for example, would require
some fine-tuning among the diagonal elements, which we do not consider acceptable.
Note that the situation deteriorates very quickly as 
$|l_{\alpha}-l_{\beta}|$ grows, due to the Gaussian dependency of the off-diagonal
coefficients. Therefore, the first issue we will address is how small can we make 
$|l_{\alpha}-l_{\beta}|$ without running into contradictions with flavour 
violating charged lepton decays. 

From the previous sections, we learn that $|l_{e}-l_{\mu}|\gtrsim 3\mu^{-1}$,
while $|l_{\tau}-l_{e,\mu}|\gtrsim 1\mu^{-1}$. Note that the lower limits 
cannot be simultaneously saturated due to the (much more stringent) $\mu\rightarrow
e\gamma$ bounds and the requirement that the charged lepton mass hierarchy is
explained within the AS scenario. For example, if we require $|l_{\tau}-l_{\mu}|
\sim 1\mu^{-1}$, in one extra dimension, $|l_{\tau}-l_{e}|\gtrsim 7\mu^{-1}$, while 
$|l_{e}-l_{\mu}|\gtrsim 8\mu^{-1}$. In two or more extra dimensions, it is possible
to obtain $|l_{\tau}-l_{e,}|\sim 2.5\mu^{-1}$ while keeping 
$|l_{e}-l_{\mu}|\gtrsim 3\mu^{-1}$ (in which case, 
$|l_{e}-l_{\mu}|$ is very close to its lower bound). It should be kept in mind
that this situation is very ``special,'' and the branching ratios for
$\mu\rightarrow eee$ and $\tau\rightarrow (e,\mu)\gamma$ are very close to the 
current experimental bounds \cite{PDG}. 

Therefore, the ``best'' possible scenario seems to be: 
the $e\mu$-element is suppressed
by at least $e^{-0.5\times 3^2}\simeq 0.01$ with respect to the diagonal 
elements, while one of the other off-diagonal elements
(we choose it to be the $\mu\tau$-element) 
can be suppressed by only $e^{-0.5\times 1^2}\sim 0.6$ with respect to the diagonal
terms. The remaining off-diagonal element 
(we will choose it to be the $e\tau$-element) is suppressed by 0.05. 
At this point, one may ask whether it is possible to find, without carefully 
choosing the order one $\alpha_{\beta\gamma}$ coefficients, phenomenologically 
viable neutrino masses and mixing angles.

In order to answer this question, we perform the following ``search.'' We 
parametrise the neutrino Majorana mass matrix by
\begin{equation}
M_{\nu}=m_{\rm Maj}
\left(\matrix{\alpha_{ee} & 0
 & \epsilon \cr 
0 & \alpha_{\mu\mu} & 
\alpha_{\mu\tau}\times 0.5 \cr 
\epsilon & 
\alpha_{\mu\tau}\times 0.5
& \alpha_{\tau\tau}}\right),
\label{scan_anarch}
\end{equation}    
where we randomly choose values for $\alpha_{(ee),(\mu\mu),(\tau\tau),(\mu\tau)}$
between 0.3 and 3.0,\footnote{we are neglecting, for simplicity, possibly large
phases associated with the order one higher dimensional Yukawa couplings
\cite{BdGR}.} while randomly choosing values for $\epsilon$ between 0 and
0.1. The $e\mu$-entries have been set to zero for simplicity. We have verified that
the results we obtain are not changed in any significant way if $e\mu$-elements
of order $0.01$ are also included. Note that we still have one free parameter, namely
$m_{\rm Maj}$. The quantity $m_{\rm Maj}\cdot\alpha_{ee}$ 
is directly constrained by searches for
neutrinoless double beta decay, which currently requires $|M_{\nu}^{ee}|\lesssim
0.35$~eV \cite{PDG,double_beta}. In our analysis, we will circumvent the dependency on 
$m_{\rm Maj}$ by computing the ratio of the mass-squared differences 
\begin{equation}
ratio\equiv\frac{\Delta m^2_{\odot}}{\Delta m^2_{\rm atm}},
\end{equation}
where $\Delta m^2$'s and mixing matrix elements are defined as in Eqs.~(\ref{window}) 
(see also the remarks which follow the equations for a comment on inverted 
hierarchies). 

\begin{figure}
\centerline{
    \psfig{file=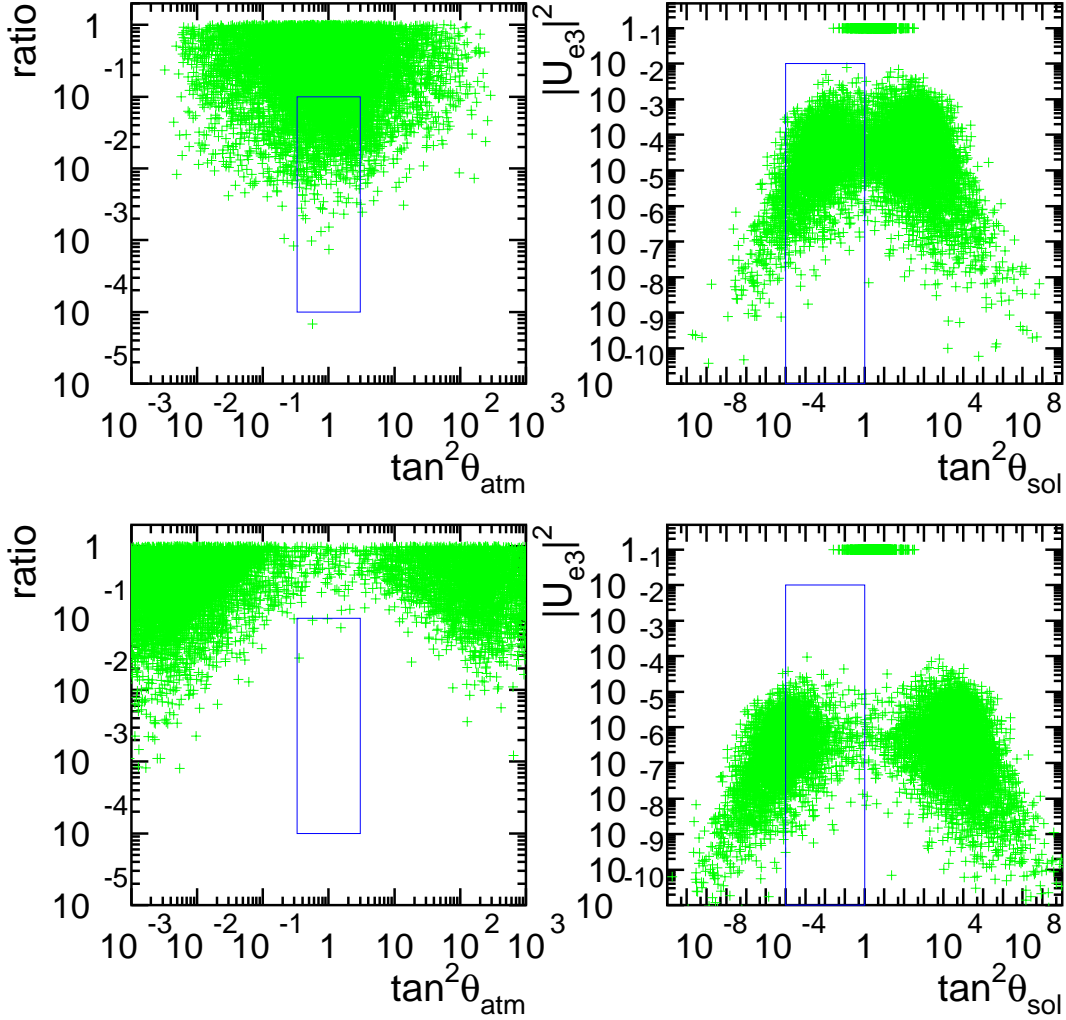,width=1.0\columnwidth}}
    \caption{Neutrino mixing parameters obtained for 10001 randomly generated
Majorana mass matrices, imposing constraints from the absence of flavour violating
charged lepton decays (see text) displayed in the following manner: 
$ratio\equiv \Delta m^2_{\odot}/\Delta m^2_{\rm atm}\times 
\tan^2\theta_{\rm atm} (left)$ and $|U_{e3}|^2\times \tan^2\theta_{\odot}$ (right).
In the top, the branching ratios for flavour changing lepton decays are close to
the current experimental bounds \cite{PDG}, while
the bottom figures explore the behaviour of the oscillation parameters as the 
prediction for the branching ratios for flavour changing charged lepton decays 
decreases (see text for details).}
    \label{random}
\end{figure}

Figs.~\ref{random}(top) depict the results obtained for 10001 randomly 
selected points. Fig.~\ref{random}(top,left) depicts the obtained values of 
$ratio \times \tan^2\theta_{\rm atm}$, while Fig.~\ref{random}(top,right) depicts 
the obtained values of $|U_{e3}|^2 \times \tan^2\theta_{\odot}$. As one could have 
expected, the atmospheric angle is generically large \cite{anarchy,minimalistic} even
though the ${\mu\tau}$-element of Eq.~(\ref{scan_anarch})
is typically suppressed by a factor of two with respect to the
diagonal elements, and $ratio$ is indeed usually of order one. 
Furthermore, $|U_{e3}|^2$ and $\tan^2\theta_{\odot}$ are usually small, even though 
large values of $\tan^2\theta_{\odot}$ can be obtained.\footnote{Note that there 
are some points for which $|U_{e3}|^2\simeq 1$. These are expected, and correspond to 
cases when $\alpha_{ee}$ is bigger than the other coefficients. In this case, the 
largest eigenvalue is proportional to $\alpha_{ee}$ and has a very large 
electron-type component. All these points are incompatible with the current 
neutrino data, of course.} 

More significant than the figures, perhaps, are the 
following numerics: 51\% of the points yield $1/3<\tan^2\theta_{\rm atm}<3$,
while 22\% (1.5\%) of the points yield $ratio<10^{-1}$ ($10^{-2}$), and 
38\% of the points yield $10^{-5}<\tan^2_{\odot}<1$ (note that the values of
$ratio$ we obtain are not small enough to accommodate the  LOW of VAC solutions 
to the solar neutrino puzzle). Combining all constraints, 1.5\% satisfy
$1/3<\tan^2\theta_{\rm atm}<3$, $10^{-1}<ratio<10^{-3}$, $0.1<\tan^2_{\odot}<1$, 
$|U_{e3}|^2<0.1$ (LMA-like solution) and 0.1\% satisfy 
$1/3<\tan^2\theta_{\rm atm}<3$, $10^{-2}<ratio<10^{-4}$, $10^{-5}<\tan^2_{\odot}<
2\times 10^{-3}$, $|U_{e3}|^2<0.1$ (SMA-like solution). 
 
What do these numbers imply? While most points do not yield 
phenomenologically acceptable parameters, there is a slightly more than 1\% chance
for obtaining the LMA solution with completely random order one Yukawa couplings.
We do not consider this too ``unlikely,'' but leave definitive conclusions 
to the reader. One concrete statement, however, is that
obtaining the SMA solution under these conditions is certainly less
likely (15 times). The reason for this is that the SMA solution requires a larger
hierarchy for the mass-squared differences, which is very hard to obtain with
order one random Yukawa couplings.  

What happens as the separation between left-handed lepton fields increases? 
Fig.~\ref{random}(bottom, left) depicts the obtained values of 
$ratio \times \tan^2\theta_{\rm atm}$ if the same procedure is repeated, except
that the suppression factor for the $\mu\tau$-element is increased by a factor of 10
(from 0.5 to 0.05). The results are rather striking. Unlike the previous case,
depicted in Fig~\ref{random}(top, left), large atmospheric mixing becomes more
``rare'' (11\% of the points yield $1/3<\tan^2\theta_{\rm atm}<3$), especially 
when $ratio$ is required to be small (0.04\% percent (none) of the cases satisfy 
$1/3<\tan^2\theta_{\rm atm}<3$ and $ratio<0.1$ (0.01)!). Therefore, in light
of the observed hierarchy of neutrino mass-squared differences, large mixing in the
atmospheric sector becomes exceedingly unlikely as the separation between $l_{\mu}$
and $l_{\tau}$ increases.

On the other hand, Fig.~\ref{random}(bottom, right)   
depicts the obtained values of $|U_{e3}|^2 \times \tan^2\theta_{\odot}$
when the original procedure is again performed, but restraining $\epsilon<0.01$.
When compared to Fig.~\ref{random}(top, right), it is easy to see that large mixing
in the solar sector happens for a smaller fraction of all randomly generated matrices.
The number of potentially good candidates is still reasonable: 28\% of all matrices
yield $10^{-5}<\tan^2_{\odot}<1$. Here, however, the probability of obtaining
LMA-like solutions (see discussion related to Fig.~\ref{random}(top)) decreases
(0.3\% compared to 1.5\%), while the number of SMA-like solutions remains the same. 
Of course, this behaviour is expected, and the situation 
deteriorates very rapidly as the upper bound on $\epsilon$ decreases. 

In summary, if neutrinos have a naturally small Majorana mass matrix, the AS
solution to flavour changing charged lepton decays and the hierarchy of charged
lepton masses naturally predicts that the neutrino masses are not strongly hierarchical
and that mixing angles are rather small. In order to obtain large enough atmospheric
mixing angles, we were forced to live very close to the current experimental bounds
imposed by the searches for charged lepton flavour violation.
We dare not make any precise predictions given the qualitative nature of
our estimates, but if flavour violating $\mu$ and (especially) $\tau$ decays 
are not observed by the next generation of experiments \cite{new_expts}, the scenario
discussed in this section will be severely constrained, if not entirely ruled out.

This is to be contrasted to the study in the previous
section, where neutrino mixing angles were also generically 
predicted to be small, while the masses were hierarchical. In order to 
obtain large mixing angles, we were forced to impose specific positions for 
the different leptonic fields. All solutions we found were either immediately
ruled out by searches for lepton flavour violating processes, or predicted that 
their branching ratios were severely suppressed, out of reach of any foreseeable
experiment. The price paid for these solutions was having to finely tune 
{\it a priori}\/ unrelated distances to more than 1 part in $10^2$.     

\setcounter{equation}{0}
\setcounter{footnote}{0}
\section{Summary and Conclusions}

We have analysed the issue of charged lepton flavour violation and neutrino masses in the
fat-brane paradigm proposed by Arkani-Hamed and Schmaltz (AS scenario). The AS
scenario is particularly suited for explaining the absence of lepton flavour 
violating muon and tau decays, and can also naturally explain why 
neutrinos are more than ten orders of magnitude lighter than the top quark, 
if right-handed neutrinos are introduced {\sl in the brane}. 

We found that, indeed, the branching ratios for flavour violating 
muon and tau decays can be very easily suppressed to levels below the current 
experimental bounds, and that very small 
neutrino masses can be obtained for separations of order $10\mu^{-1}$. 
The AS scenario, however, seems to have a harder time accounting for
the large neutrino mixing which has been observed in the atmospheric neutrino data. 
This is a peculiar feature of the AS scenario for fermion masses: it naturally
accommodates large mass hierarchies and small mixing angles, while it seems to 
require additional ``structure'' 
({\it e.g.,}\/ different pairs of fields have to be separated by very similar
distances) in order to explain large mixing angles.

We studied the case of two lepton families in detail. The situation
seems ``finely tuned'' when the the bounds imposed by charged lepton flavour
violation are also taken into account. We showed that the charged lepton
mass matrix is constrained to be
quasi-diagonal, and that, in order to obtain large atmospheric
mixing and the appropriate $\Delta m^2$, distances which are {\it a priori}\/
unrelated are required to agree with one another better than one part in 100.
We proceeded to analyse the case of three lepton flavours, and while we found
that all the different solutions to the solar neutrino puzzle can be accommodated
while keeping the absolute value of the 
$U_{e3}$ element of the leptonic mixing matrix small, we
argued that the amount of fine-tuning required to simultaneously satisfy all neutrino 
data is ``larger'' than in the two-family case.    

We then proceeded to discuss the effect of explaining the absence of flavour changing 
tau and muon decays and charged lepton masses in the AS scenario on theories where small
Majorana neutrino masses are generated by breaking lepton number in a far away brane.
In this case, in order to obtain large mixing in the atmospheric sector, we were forced
to ``approximate'' different lepton doublet fields to the point that the current
experimental upper bounds on the branching ratios of some rare tau and muon decays
were almost saturated. Therefore, we expect these rare decays to be observed in the 
next round of experiments, if such a scenario were indeed realised in nature. 
Furthermore, very hierarchical neutrino mass-squared differences were not 
attainable, meaning that the solar neutrino puzzle would have to be solved either 
by the SMA or the LMA  solutions.

In conclusion, we would like to emphasise that, in spite of the fine-tuning problems
we highlighted in Sec.~3, the AS scenario should be regarded as a novel mechanism
for understanding small mixing angles, which does not necessarily require the
existence of fat branes or even large extra dimensions. Similar results are obtained
in models with many different ``thin'' branes which are separated in the 
extra dimensions. Furthermore, if the extra dimensions are not large 
(and the hierarchy problem is solved by some other means), our results for 
neutrino masses and mixing angles would still apply, except for the flavour 
changing constraints from charged lepton decays, which would be absent. 
Finally, the fine-tuning 
which we pointed out could (and should, in our opinion) be interpreted
as a challenge to be faced by candidates for the localising mechanism and/or 
the structure of the compact dimensions.

\section*{Acknowledgements}

AdG would like to thank Martin Schmaltz for some useful conversations and the Aspen 
Center for Theoretical Physics, where some of the ideas presented here were first
considered. GCB and MNR thank the CERN Theory Division for its hospitality and
Funda\c{c}\~ao para Ci\^encia e Tecnologia (Portugal) for partial support through
Project POCTI/1999/FIS/36288, and the European Comission for partial support 
under the RTN contract HPRN-CT-2000-00149.

\appendix

\setcounter{equation}{0}
\setcounter{footnote}{0}
\section{Numerical Scan of the 3 Generation Parameter Space}

\begin{figure}
\centerline{
    \psfig{file=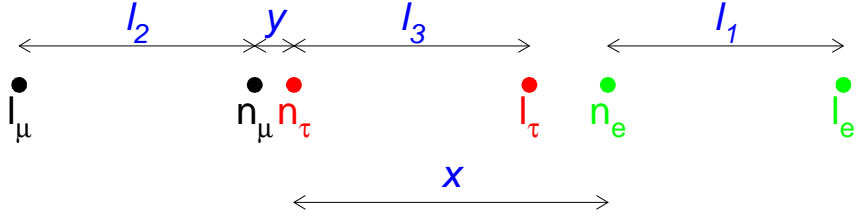,width=1.0\columnwidth}}
    \caption{General three family configuration in one extra dimension. There are
five independent distances, which are chosen to be $l_1,l_2,l_3,x,y$. Note that
in order to span the entire location space, four out of the five parameters should
take positive and negative values, while the sign of the fifth one remains fixed.}
    \label{3gen_draw}
\end{figure}

In this appendix, we describe a numerical scan over the three generation, one
extra dimension location space of the lepton doublets and right-handed neutrino
singlets. In this case, specifying five distances 
suffices to determine the relative distance between the six different fermion 
fields. 
These distances are pictorially defined in Fig.~\ref{3gen_draw}, and are:
$l_1$, $l_2$, $l_3$, the distances between $l_{\alpha}$ and $n_{\alpha}$, 
$\alpha=e,\mu,\tau$; $x$, the distance between $n_{\tau}$ and $n_{e}$; 
$y$, the distance between $n_{\tau}$ and $n_{\mu}$. Note that the sign of four 
out of the five distances is also meaningful 
(for example, when $y<0$, $n_{\mu}$ is ``to the right'' of $n_{\tau}$). 

In the scan, we varied $l_1$ ($l_2$) from $-20\mu^{-1}$ to $20\mu^{-1}$ 
($-20.1\mu^{-1}$ to $20.1\mu^{-1}$ ) in steps
of 0.2, $x$ and $y$ from $-20\mu^{-1}$ to $20\mu^{-1}$ in steps of 0.4, and
$l_3$ from $0.1\mu^{-1}$ to $20.1\mu^{-1}$ in steps of 0.2. 
Out of all the points, we only
consider these which yield neutrino mass 
matrices such that $(m_{\nu}m_{\nu}^{\dagger})_{ij}<10$~eV$^2$ for all 
$i,j=e,\mu,\tau$.
This is done in order to guarantee that all neutrino masses squared are roughly 
less than 10~eV$^2$. One consequence of this constraint is that for all $l_i$,
$|l_i|\gtrsim 7.0$, while certain bands for $x$ and $y$ are excluded. 
 
Of the remaining matrices, we identify the ones whose eigenvalues
and eigenvectors fall in the following window:
\begin{eqnarray}
&|U_{e3}|^2<0.1, \nonumber \\
&1/3<(|U_{\mu3}|/|U_{\tau3}|)^2\equiv\tan^2\theta_{\rm atm}<3, \nonumber \\
&10^{-5}<(|U_{e2}|/|U_{e1}|)^2\equiv\tan^2\theta_{\odot}<10, \nonumber \\
&10^{-3}~{\rm eV^2}<(m_3^2-m_2^2)\equiv\Delta m^2_{\rm atm}<10^{-2}~{\rm eV^2}, 
\nonumber \\
&10^{-10}~{\rm eV^2}<(m_2^2-m_1^2)\equiv\Delta m^2_{\odot}<10^{-4}~{\rm eV^2}, 
\label{window}
\end{eqnarray}
where we define the neutrino mixing matrix as 
$\nu_i=U_{\alpha i}\nu_{\alpha}$, where $\nu_{\alpha}$, 
$\alpha=e,\mu,\tau$ are weak eigenstates and $\nu_i$, $i=1,2,3$ are mass eigenstates,
with masses $m_i$. The eigenstates are organised in ascending order of mass-squared.
When $m_3^2-m_2^2>m_2^2-m_1^2$ (normal hierarchy), 
the window defined above contains the region of
the parameter space which satisfies all of the experimental constraints 
\cite{neutrinos}. If
$m_3^2-m_2^2<m_2^2-m_1^2$ (inverted hierarchy) the same window can be used after
relabelling the mass eigenstates $1\rightarrow 3\rightarrow 2\rightarrow 1$, and
defining $(m_2^2-m_3^2)\equiv\Delta m^2_{\rm atm}$. It should be noted 
that during the numerical scan we did not find any matrix which yields an inverted 
mass hierarchy and is consistent with the constraints imposed above.

The results of the scan are best described in words. Out of roughly $2\times 10^{9}$
initially selected matrices, only 440 fall within the window defined above. 
The events are rejected at the following rate: 
99\% fail the mass constraints. From the surviving points, 96\% fail to satisfy 
the $\tan^2\theta_{\odot}$ constraint. From the remaining points, 97\% fail the 
$\tan^2\theta_{\rm atm}$ constraint, while 
98\% of the remainder fall out of the $|U_{e3}|^2$ bound. Note that only one out
of roughly 5 million events survives all the constraints we define above.

Many comments are in order. First, in should be readily noted that the two ``angle 
constraints'' combine to remove 99.9\% of the candidate matrices, 
ten times more than the ``mass constraints.''
Second, it should be pointed out that the mass constraints and the angle constraints
``commute'', in the sense that, independent of the order in which they are applied, 
the percentage of the points which is removed is the same. This implies that there
is no correlation between the values of the masses and the values of the mixing
angles. The same is not true
for the $\tan^2\theta_{\odot}$ and $\tan^2\theta_{\rm atm}$ constraints. 
Third,
the $|U_{e3}|^2$ constraint is highly correlated with the angle constraints. If
applied first, the $|U_{e3}|^2$ constraint only cuts a small fraction of all the
matrices, while if applied last, it removes a very significant fraction of
the events which pass the angle constraints. Therefore, events which have a small
$|U_{e3}|$ generically have very small mixing in the solar and/or the atmospheric
sector. This can be understood from the qualitative discussion included in the
beginning of Sec.~3: the AS scenario prefers small mixing angles, and a 
generic set of points will yield a mixing matrix which has only ``0s'' and 
``1s'' (one 1 for each column and row). Therefore, if we impose $|U_{e3}|^2$ small,
the generic matrices which are left have some other $U_{\alpha3}$ and $U_{ei}$ 
which is close to one, which necessarily means that 
$\tan^2\theta_{\odot, \rm atm}$ are either very large or very small.
   
The selected points yield a varied spectrum of solar angles and masses-squared,
and no value of $\Delta m^2_{\odot}$ and $\tan^2\theta_{\odot}$ is strongly 
preferred. All points, without exception, choose the second and third generation
fields to fall into either the configuration depicted in Fig.~\ref{geoms}(A) or
Fig.~\ref{geoms}(B) (note that the configurations where $n_{\mu}$ and 
$n_{\tau}$ are exchanged are, of course, also present in equal numbers), 
while $l_1$ and $n_1$ are placed in various locations.
Furthermore, the choices for $l$ and $\delta$ in the case of
the configuration Fig.~\ref{geoms}(A), and $l_{21}$ and $l_{22}$ in the case
of configuration Fig.~\ref{geoms}(B) are exactly the ones obtained in the last
line of Tables~\ref{table1} and \ref{table2}, respectively. This implies,
as one would naively expect, that in order to obtain large mixing in the 
``atmospheric sector'' the same tuning of parameters as the ones discussed in
Sec.~3.1 is required. 

\begin{figure}
\centerline{
    \psfig{file=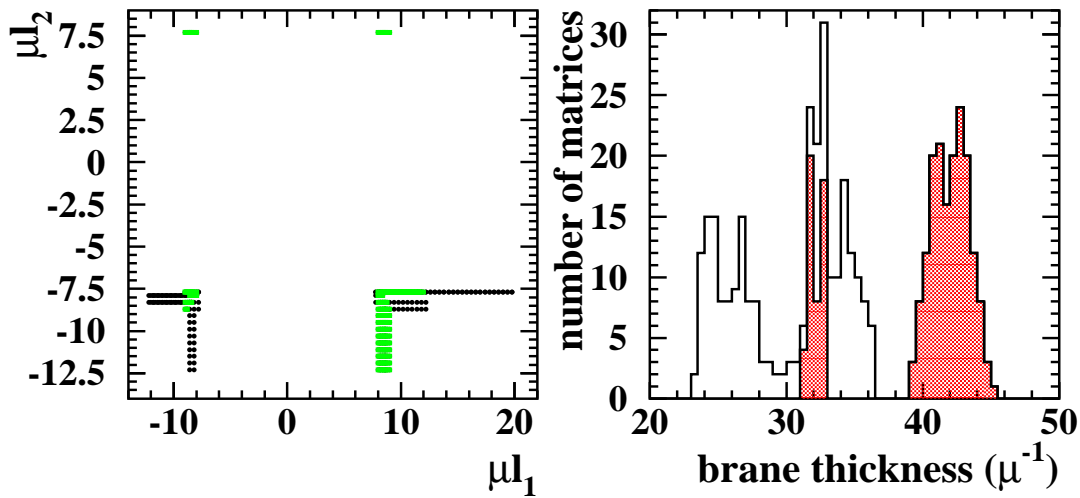,width=1.0\columnwidth}}
    \caption{Scatter plot of $l_2\times l_1$ of all the 440 points which
satisfy all constraints imposed by the neutrino data (left, see text). The light 
[green] points also satisfy the ``flavour changing constraints,''  
$|l_{\tau}-l_{e,\mu}| > 1\mu^{-1}$ and $|l_{\mu}-l_e| > 3\mu^{-1}$. 
Histogram of the brane-thickness for the same 440 points (right). 
The painted histogram corresponds
to events which satisfy the flavour changing constraints.}
    \label{scan_res}
\end{figure}

Fig.~\ref{scan_res}(left) is a scatter plot of the values of $l_2$ and $l_1$
of the points which pass all of the constraints mentioned above. The light [green] 
points are the ones which also pass the constraints that two lepton doublets are
not too close to one another (flavour changing constraints): 
$|l_{\tau}-l_{e,\mu}|\gtrsim 1\mu^{-1}$ and $|l_{\mu}-l_e|\gtrsim 3\mu^{-1}$ (see 
Fig.~\ref{fig:mu-egamma}).
This constraint
is sufficient for satisfying all flavour changing constraints described in Sec.~2
and also guarantee that the mixing in the charged lepton sector is negligible.
218 out of the 440 points satisfy this extra requirement. 

Fig.~\ref{scan_res}(right) is a histogram of the largest distance between two
lepton fields, and can be interpreted as a lower bound on the brane ``thickness.''
The solid painted histogram is for points which satisfy 
the flavour changing constraints. Note that only
``very thick'' branes are obtained, especially for matrices which do not have 
two $l$s on top of each other. This could be the subject of serious concern, since
branes should not be too ``fat'' if one wants to solve the gauge hierarchy problem
and still have the gauge bosons propagate freely in the entire fat-brane (see 
\cite{MS} for details). We prefer, instead, to view this as an indication that one
extra dimension seems to be problematic, and that if one considers branes which
are fat in more dimensions similar configurations can be obtained for much ``thiner''
branes. This is very easy to see. All one has to do is rotate a number of lepton
fields, keeping the ``smaller'' distances fixed, while making sure that the
``larger'' distances remain large. Because of the Gaussians dependency of the
Yukawa couplings on the distances, these configurations will be, in practice,
indistinguishable from the one-dimensional ones we discuss here.  

One may worry whether configurations Fig.~\ref{geoms}(C) and Fig.~\ref{geoms}(D)
can yield proper solar parameters, since these configurations never show up
in our general scan. In order to check this, we numerically searched for solutions
by starting from configurations Fig.~\ref{geoms}(C) and Fig.~\ref{geoms}(D), choosing
distances such that the atmospheric neutrino puzzled is solved and varying the
parameters $l_1$ and $x$ (see Fig.~\ref{3gen_draw}). Indeed, we are able to find a 
large number of solutions as long as $l_{\mu}$ is not on top of $l_{\tau}$, but
extremely close. This is easy to understand: when $l_{\mu}$ is exactly on top of
$l_{\tau}$, the neutrino mass matrix has a zero eigenvalue, whose eigenvector is
very close to $1/\sqrt{2}(|\nu_{\mu}\rangle-|\nu_{\tau}\rangle)$. This implies
that the $|\nu_e\rangle$ state is contained in the two heaviest mass eigenstates,
which makes it very hard to solve the solar neutrino puzzle. The only alternative is to
choose the other distances such that an inverted mass hierarchy is obtained, but
such a situation is extremely finely-tuned.\footnote{One may wonder whether an SMA-like 
solution to the solar neutrino puzzle cannot be obtained by varying the
parameters slightly. This is not the case since $\nu_e$ is going to be 
predominately {\sl heavy} and even though $\sin^2 2\theta_{\odot}\ll 1$ this point
belongs to the ``dark side'' \cite{dark_side} of the parameter space, 
$\tan^2\theta_{\odot}\gg 1$.} It should be noted that all solutions found
starting from the configurations \ref{geoms}(C) and (D) are ruled out by 
flavour changing constraints. Indeed, as was pointed out in Sec.~3.1, 
the configuration Fig.~\ref{geoms}(D) is already rule out in the
case of two generations, while satisfactory solutions with the configuration
Fig.~\ref{geoms}(C) are extremely finely tuned.

Another worry is whether much thinner branes cannot be obtained. One possibility,
for example, is to have all right-handed neutrinos very close to each other. 
Since we are restraining ourselves
to only one dimension, these configurations are very hard to obtain, and the
flavour changing constraints make them even harder to find. Nonetheless, we have
been able to find (numerically) examples which fit this description and do not 
yield too large charged lepton mixing or flavour changing effects.

\end{document}